\documentclass[english,aps,superscriptaddress]{revtex4-1}
\usepackage[T1]{fontenc}
\usepackage[latin9]{inputenc}
\usepackage{color}
\usepackage{graphicx}
\usepackage{babel}
\usepackage{bm}
\usepackage{amsmath}
\usepackage{lineno}
\usepackage{hyperref}

\makeatletter

\def\gsim{\mathrel{\rlap{\lower4pt\hbox{\hskip1pt$\sim$}}
    \raise1pt\hbox{$>$}}}         %greater than or approx. symbol
\def\lsim{\mathrel{\rlap{\lower4pt\hbox{\hskip1pt$\sim$}}
    \raise1pt\hbox{$<$}}}         %less than or approx. symbol

 \@ifundefined{textcolor}{}
 {%
   \definecolor{BLACK}{gray}{0}
   \definecolor{WHITE}{gray}{1}
   \definecolor{RED}{rgb}{1,0,0}
   \definecolor{GREEN}{rgb}{0,1,0}
   \definecolor{BLUE}{rgb}{0,0,1}
   \definecolor{CYAN}{cmyk}{1,0,0,0}
   \definecolor{MAGENTA}{cmyk}{0,1,0,0}
   \definecolor{YELLOW}{cmyk}{0,0,1,0}
 }
\makeatother
\newcommand{\lp}{\left(}
\newcommand{\rp}{\right)}

\begin{document}
\vspace{2cm} DESY-17-015
%\linenumbers

\title{The photon PDF from high-mass Drell Yan data at the LHC}

%\author{F.~Giuli, xFitter Developers' team: V.~Bertone, D.~Britzger, S.~Carrazza, A.~Cooper-Sarkar, A.~Glazov, K.~Lohwasser, A.~Luszczak,  F.~Olness,
% R.~Placakyte, V.~Radescu,  J.~Rojo, R.~Sadykov, P.~Shvydkin,  O.~Zenaiev;  and M.~Lisovyi}

\author{F.~Giuli}
\affiliation{University of Oxford,1 Keble Road, Oxford OX1 3NP, United Kingdom}
%\author{and the xFitter Developers team:}
%\noaffiliation
\author{and the xFitter Developers' team: V.~Bertone}
\affiliation{Department of Physics and Astronomy,  VU University, NL-1081 HV Amsterdam, \\ and Nikhef Theory Group Science Park 105, 1098 XG Amsterdam, The Netherlands}
\author{D.~Britzger}
\affiliation{DESY Hamburg, Notkestrasse 85 D--22609, Hamburg, Germany}
\author{S.~Carrazza}
\affiliation{CERN, CH--1211 Geneva 23, Switzerland}
\author{A.~Cooper-Sarkar}
\affiliation{University of Oxford,1 Keble Road, Oxford OX1 3NP, United Kingdom}
\author{A.~Glazov}
\affiliation{DESY Hamburg, Notkestrasse 85 D--22609, Hamburg, Germany}
\author{K.~Lohwasser}
\affiliation{DESY Zeuthen, Platanenallee 6 D--15738, Zeuthen, Germany}
\author{A.~Luszczak}
\affiliation{T.Kosciuszko Cracow University of Technology, 30-084 Cracow, Poland}
\author{F.~Olness}
\affiliation{SMU Physics, Box 0175 Dallas, TX  75275-0175, United States of America}
\author{R.~Pla\v cakyt\.e}
\affiliation{Institut f\"ur Theoretische Physik, Universit\"at Hamburg \\ Luruper Chaussee 149, D--22761 Hamburg, Germany}
\author{V.~Radescu}
\affiliation{University of Oxford,1 Keble Road, Oxford OX1 3NP, United Kingdom}
\affiliation{CERN, CH--1211 Geneva 23, Switzerland}
\author{J.~Rojo}
\affiliation{Department of Physics and Astronomy,  VU University, NL-1081 HV Amsterdam, \\ and Nikhef Theory Group Science Park 105, 1098 XG Amsterdam, The Netherlands}
\author{R.~Sadykov}
\author{P.~Shvydkin}
\affiliation{Joint Institute for Nuclear Research (JINR), Joliot-Curie 6, 141980, Dubna, Moscow Region, Russia}
\author{O.~Zenaiev}
\affiliation{DESY Hamburg, Notkestrasse 85 D--22609, Hamburg, Germany}
\author{M.~Lisovyi}
\affiliation{Physikalisches Institut, Ruprecht-Karls-Universit\"at Heidelberg, Heidelberg, Germany}

\date{\today}
\begin{abstract}
  Achieving the highest precision for theoretical predictions
  at the LHC requires the 
  calculation of hard-scattering cross-sections that include
  perturbative QCD corrections up to (N)NNLO and electroweak (EW)
  corrections up to NLO.
  Parton distribution functions (PDFs) need to be
  provided with matching accuracy, which in the case of QED effects
  involves introducing the photon parton distribution of the proton,
  $x\gamma(x,Q^2)$.
  In this work a determination of the photon PDF from
  fits to recent ATLAS measurements of high-mass Drell-Yan dilepton
  production at $\sqrt{s}=8$ TeV is presented.
  This analysis is based on the {\tt xFitter} framework,
  and has required improvements both in the {\tt APFEL} program, to account
  for NLO QED effects, and in the {\tt aMCfast} interface to account
  for the photon-initiated contributions in the EW calculations within
  {\tt MadGraph5\_aMC@NLO}.
  The results are compared with other recent QED fits and
  determinations of the photon PDF, consistent results are found.
%  finding in particular agreement with the LUXqed
%  and the HKR calculations within uncertainties.
% for the
%  kinematical range where the ATLAS data has sensitivity to
%  $x\gamma(x,Q)$.
\end{abstract}
\maketitle

\tableofcontents{}

%%%%%%%%%%%%%%%%%%%%%%%%%%%%%%%%%%%%%%%%%%%%%%%%%%
\section{Introduction}

Precision phenomenology at the LHC requires theoretical calculations
which include not only QCD corrections, where NNLO is rapidly becoming
the standard, but also electroweak (EW) corrections, which are
particularly significant for observables directly sensitive to the TeV
region, where EW Sudakov logarithms are enhanced.
An important ingredient of these electroweak corrections is the photon
parton distribution function (PDF) of the proton, $x\gamma(x,Q^2)$, which must be introduced to absorb
the collinear divergences arising in initial-state QED emissions.

The first PDF fit to include both QED corrections and a photon PDF was
MRST2004QED~\cite{Martin:2004dh}, where the photon PDF was taken from
a model and tested on HERA data for direct photon production.
Almost 10 years later, the NNPDF2.3QED analysis~\cite{Ball:2012cx,Ball:2013hta} provided a
first model-independent determination of the photon PDF based on
Drell-Yan (DY) data from the LHC.
The resulting photon PDF was however affected by large uncertainties
due to the limited sensitivity of the data used as input to that fit.
The determination of $x\gamma(x,Q^2)$ from NNPDF2.3QED was later combined
with the state-of-the-art quark and gluon
PDFs from NNPDF3.0, together with an improved QED evolution,
to construct the NNPDF3.0QED set~\cite{Bertone:2016ume,Ball:2014uwa}.
The CT group has also  released a QED fit using a similar
strategy as the MRST2004QED one, named the CT14QED set~\cite{Schmidt:2015zda}.

A recent breakthrough concerning the determination of the photon content of
the proton
has been the realization that  $x\gamma(x,Q^2)$
can be calculated in terms of
inclusive lepton-proton deep-inelastic scattering (DIS) structure functions.
The photon PDF resulting from this
strategy is called  LUXqed~\cite{Manohar:2016nzj} and its residual uncertainties are now at the few
percent level, not too different from those
of the quark and gluon PDFs.
A related approach by the HKR~\cite{Harland-Lang:2016apc}
group, denoted by HKR16 in the following, also leads to a similar photon PDF
as compared to the LUXqed calculation, although in this case no estimate
for the associated uncertainties is provided.

The aim of this work is to perform a direct determination of the
photon PDF from recent high-mass Drell-Yan measurements from
ATLAS  at $\sqrt{s}=8$ TeV~\cite{Aad:2016zzw}, and to
compare it with some of the existing determinations of $x\gamma(x,Q^2)$
mentioned above.
Note that
earlier measurements of high-mass DY from ATLAS and CMS were presented
in Refs.~\cite{CMS:2014jea,Chatrchyan:2013tia,Aad:2013iua}.
The ATLAS 8 TeV DY data are provided in terms of both
single-differential cross-section distributions in the dilepton invariant mass,
$m_{ll}$, and of double-differential 
cross-section distributions in $m_{ll}$ and $|y_{ll}|$, the absolute value of rapidity of the
lepton pair, and in $m_{ll}$ and $\Delta\eta_{ll}$, the difference in
pseudo-rapidity between the two leptons.
Using the Bayesian reweighting method~\cite{Ball:2011gg,Ball:2010gb}
applied to NNPDF2.3QED, it was shown in the
same publication~\cite{Aad:2016zzw} that these
measurements provided significant information on $x\gamma(x,Q^2)$.

The goal of this study is therefore to investigate further these
constraints from the ATLAS high-mass DY measurements on the photon PDF,
this time by means of a direct PDF fit performed within the
open-source {\tt xFitter} framework~\cite{Alekhin:2014irh}.
State-of-the-art theoretical calculations are employed, in particular
the inclusion of  NNLO QCD and NLO QED corrections to the PDF evolution and
the computation of the DIS structure
functions as implemented in the {\tt APFEL} program~\cite{Bertone:2013vaa}.
The implementation of NLO QED effects in {\tt APFEL} is
presented here for
the first time.
The inclusion of NLO QED evolution effects is cross checked using the independent
{\tt QEDEVOL} code \cite{Sadykov:2014aua} based on the {\tt QCDNUM} evolution program \cite{Botje:2010ay}.

The resulting determination of $x\gamma(x,Q^2)$
represents an important validation test of
recent developments in theory and data concerning
our understanding of the nature and implications
of the photon PDF.

The outline of this paper is as follows.
Sect.~\ref{sec:theory} reviews the ATLAS 8 TeV high-mass DY data together
with the theoretical formalism of the DIS and Drell-Yan cross-sections
used in the analysis.
Sect.~\ref{sec:fitsettings} presents the settings of the PDF
fit within the {\tt xFitter} framework.
The fit results are then discussed in Sect.~\ref{sec:results}, where
they are compared to determinations by other groups.
Finally, Sect.~\ref{sec:conclusions} summarises and discusses the results and
future lines of investigation.
Appendix~\ref{sec:appendixAPFEL} contains a detailed
description of the implementation and validation of NLO QED
corrections to the DGLAP PDF evolution equations
and DIS structure functions, which are
available now in {\tt APFEL}.

\section{Data and theory}
\label{sec:theory}

\begin{figure}[t]
  \begin{center}
    \includegraphics[width=17cm]{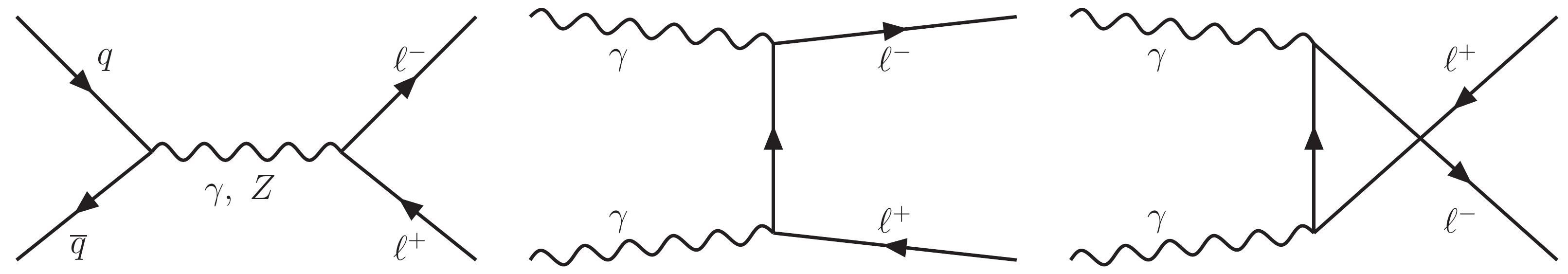}
    \end{center}
    \caption{Diagrams that contribute to lepton-pair production at
      hadron colliders at the Born level.}
\label{fig:photoninduced}
\end{figure}

In this work, the photon content of the proton $x\gamma(x,Q^2)$ is
extracted from a PDF analysis based on the combined inclusive DIS
cross-section data from HERA~\cite{Abramowicz:2015mha}
supplemented by the ATLAS measurements of high-mass Drell-Yan
differential cross sections at $\sqrt{s}=8$ TeV~\cite{Aad:2016zzw}.
The HERA inclusive data are the backbone of modern PDF fits, providing
information on the quark and gluon content of the proton, while
the high-mass Drell-Yan data provide a direct sensitivity to the
photon PDF.
As illustrated in Fig.~\ref{fig:photoninduced}, 
 dilepton production at hadron colliders can arise
from either quark-antiquark $s$-channel scattering, or from
photon-photon $t$- and $u$-channel scattering mediated by a lepton.

The ATLAS high-mass Drell-Yan 8 TeV measurements are presented in terms
of both the
single-differential (1D) invariant-mass distribution,
$d\sigma/dm_{ll}$, as well as double-differential (2D)
distributions in $m_{ll}$ and $y_{ll}$, namely
$d^{2}\sigma/dm_{ll}d|y_{ll}|$, and in $m_{ll}$ and $\Delta\eta_{ll}$,
$d^{2}\sigma/dm_{ll}\Delta\eta_{ll}$.
%, where $m_{ll}$, $y_{ll}$ and $\Delta\eta_{ll}$
%denote the invariant mass, rapidity, and separation in pseudo-rapidity
%of the lepton pair, respectively.
%
For the invariant-mass 1D distribution, there are 12 bins between $m_{ll}=116$ GeV
and 1.5 TeV; and for both double-differential distributions,
there are five different bins in invariant mass,
from the lowest bin with 116 GeV < $m_{ll}$ <
150 GeV to the highest bin with 500 GeV < $m_{ll}$ < 1500 GeV.
The first three (last two) $m_{ll}$ bins of the 2D distributions are divided into 12 (6) bins
with fixed width, extending up to 2.4 and 3.0 for the $|y_{ll}|$
and $|\Delta\eta_{ll}|$ distributions, respectively.

The photons which undergo hard scattering in the
$\gamma\gamma \to ee$  process from Fig.~\ref{fig:photoninduced}
can be produced by either  emission from the
proton as a whole (the ``elastic'' component) or radiated by the constituent quarks
(the ``inelastic'' component).
In the ATLAS high-mass DY measurement, the double elastic component is suppressed
 by the
 requirement of having more than two primary tracks~\cite{Aad:2016zzw}.
 Moreover, the inelastic-elastic photon scattering corresponds to only about
 10\% of the total measured cross sections.
 From the theory point of view, the photon PDF extracted from the
 fit is by construction the sum of the elastic and inelastic contributions,
 though this analysis is mostly sensitive to the latter.

For the calculation of NLO high-mass Drell-Yan cross sections, the
{\tt MadGraph5{\_}aMC@NLO}~\cite{Alwall:2014hca} program is used, which
includes the contribution from photon-initiated diagrams, interfaced
to {\tt APPLgrid}~\cite{Carli:2010rw} through {\tt
  aMCfast}~\cite{amcfast}.
A tailored version of {\tt APPLgrid} is used, accounting for
the contribution of the photon-initiated processes \footnote{Modified version of {\tt APPLgrid} available at: $https://github.com/scarrazza/applgridphoton$}.
The calculation is performed in the $n_f=5$ scheme neglecting mass
effects of charm and bottom quarks in the matrix elements, as
appropriate for a high-scale process.
These NLO theoretical predictions match the  
analysis cuts of the data, with $m_{ll}\ge 116$ GeV, $\eta_l\le 2.5$, and
$p_T^l \ge 40$ GeV$~(30)$ GeV for the leading (sub-leading) lepton being
the most important ones.
As discussed below, the NLO calculations are then supplemented by
NNLO/NLO $K$-factors obtained from {\tt FEWZ}~\cite{Gavin:2012sy}.
The NLO EW corrections to the DY processes are also estimated
using {\tt FEWZ}. 
The photon-initiated process is taken at LO since this corresponds to
the {\tt APPLgrid} implementation and the NLO corrections are
very small compared to the data accuracy.
%

%%%%%%%%%%%%%%%%%%%%%%%%%%%%%%%%%%%%%%%%%%%%%%%%%%%%%%%%%%%%%%%%
%%%%%%%%%%%%%%%%%%%%%%%%%%%%%%%%%%%%%%%%%%%%%%%%%%%%%%%%%%%%%%%%

The DIS structure functions and PDF evolution are computed with the {\tt
  APFEL} program~\cite{Bertone:2013vaa}, which is currently accurate
up to NNLO in QCD and NLO in QED, including the relevant mixed
QCD+QED corrections.
This means that, on top of the pure QCD
contributions, the DGLAP evolution
equations~\cite{Gribov:1972ri,Dokshitzer:1977,Altarelli:1977zs} are
solved including the $\mathcal{O}\lp \alpha_s\alpha\rp$ and
$\mathcal{O}\lp \alpha^2\rp$ corrections to the splitting functions.
%
%Concerning the DIS structure functions, 
Corrections of $\mathcal{O}\lp \alpha\rp$ are also included leading to a (weak)
explicit dependence of the predictions on the photon PDF.
Details
of the implementation of these corrections and of their numerical
impact are given in Appendix~\ref{sec:appendixAPFEL}.
Heavy-quark (charm and bottom) mass effects to DIS structure functions
are taken into account using the FONLL-B~(C) general-mass
scheme~\cite{Forte:2010ta} for the NLO~(NNLO) fits.
The numerical values of the heavy-quark masses in the mass parameter 
scheme
are taken to be $m_c=1.47~$GeV and $m_b=4.5~$GeV as determined in~\cite{Abramowicz:2015mha},  
consistent with the latest PDG averages~\cite{Agashe:2014kda}.
The reference values of the QCD and QED coupling constants are chosen to be
$\alpha_s(m_Z)=0.118$ and $\alpha(m_\tau=1.777\mbox{ GeV})=1/133.4$, again consistent
with the PDG recommended values.

In the  calculation of the Drell-Yan cross section, the
dynamical renormalisation $\mu_{R}$ and factorisation $\mu_{F}$
scales are used, which are set equal to the scale of invariant mass $m_{ll}$,
both for the quark- and gluon-induced and for the photon-induced
contributions.
The choice of other values for these scales in the QED diagrams,
such as a fixed scale $\mu_R=\mu_F=M_Z$, leads 
to  variations of the photon-initiated cross-sections of at most a few percent.
The choice of the scale for the photon PDF  is further discussed
in~\cite{Dittmaier:2009cr}.
For the kinematics of the ATLAS DY data, the ratio between the photon-initiated
contributions and  quark- and gluon-induced dilepton production is largest
for central rapidities and large invariant masses. 
For the most central (forward) rapidity
bin, $0 < |y_{ll}| < 0.2$ ($2.0 < |yll| < 2.4$), the ratio
between the QED and QCD contributions varies between 2.5\%~(2\%) at low
invariant masses and 12\%~(2.5\%) for the highest $m_{ll}$ bin.
Therefore, data from the central region will exhibit the highest
sensitivity to $x\gamma(x,Q^2)$.

The {\tt MadGraph5{\_}aMC@NLO} NLO QCD and LO QED calculations used in this work have been
benchmarked against the corresponding predictions  obtained with the {\tt FEWZ}
code~\cite{Gavin:2012sy}, finding agreement within statistical uncertainties of
the predictions for both the 1D and the 2D distributions.

In order to achieve NNLO QCD and NLO EW accuracy in our theoretical
calculations, the NLO QCD and LO QED cross-sections computed with
{\tt MadGraph5{\_}aMC@NLO}
have been supplemented by bin-by-bin $K$-factors
defined as:
\begin{equation}
  \label{eq:kfactor}
  K(m_{ll},|y_{ll}|) \equiv\frac{\rm NNLO\  QCD  + NLO\  EW}{\rm NLO\  QCD + LO\  EW} \, ,
\end{equation}
using the MMHT2014 NNLO~\cite{Harland-Lang:2014zoa} PDF set both in
the numerator and in the denominator. 
The $K$-factors have been computed using
   {\tt FEWZ}  with the same settings and analysis cuts as the corresponding
   NLO calculations of {\tt MadGraph5{\_}aMC@NLO}.
This approximation is justified since NNLO $K$-factors depend very
mildly on the input PDF set, see for example~\cite{Czakon:2016olj}.
The photon induced contribution, as provided in \cite{Aad:2016zzw},  has been explicitly 
subtracted from the {\tt FEWZ} predictions.
   Fig.~\ref{fig:kf} shows the $K$-factors of  Eq.~({\ref{eq:kfactor})
  corresponding to the double differential $(m_{ll},|y_{ll}|)$
  cross sections 
  as a function of the dilepton
  rapidity $|y_{ll}|$, where each set of points corresponds to a different
  dilepton invariant mass $m_{ll}$ bin.
  The $K$-factors vary between 0.98 and 1.04,
  highlighting the fact that higher-order corrections to the Drell-Yan
  process are moderate, in particular at low values of $m_{ll}$ and in
  the central region.
  Even at forward rapidities, the $K$-factors modify the NLO result by
  at most 4\%.

  %%%%%%%%%%%%%%%%%%%%%%%%%%%%%%%%%%%%%%%%%%%%%%%%%%%%%%%%
\begin{figure}[t]
\includegraphics[width=10cm]{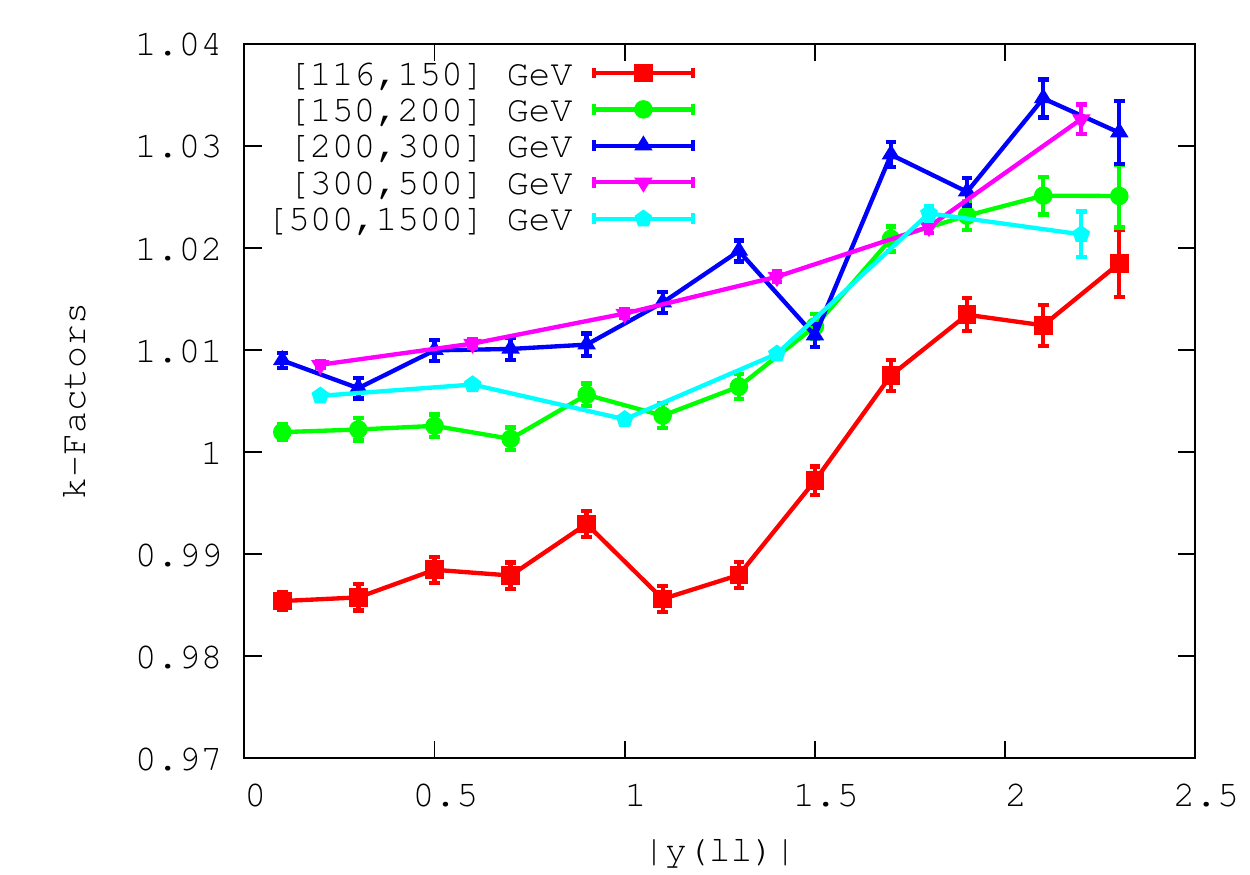}
\caption{The NNLO/NLO $K$-factors, defined in Eq.~(\ref{eq:kfactor}),
  that account for higher order QCD and EW effects to the high-mass
  Drell-Yan cross sections with the photon induced contribution subtracted,
  as a function of the dilepton rapidity $|y_{ll}|$.
  Each set of points corresponds to a different bin in the dilepton
  invariant mass $m_{ll}$.
}
\label{fig:kf}
\end{figure}
%%%%%%%%%%%%%%%%%%%%%%%%%%%%%%%%%%%%%%%%%%%%%%%%%%%%%%%%

\section{Settings}
\label{sec:fitsettings}

This section presents the settings of the PDF fits, including the details about the
parametrisation of the photon PDF $x\gamma(x,Q^2)$, which have been
carried out using the open-source {\tt xFitter}
framework~\cite{Alekhin:2014irh}.
First of all, the scale $Q_0^2$ at which PDFs are parametrised is taken to be
$Q_0^2 = 7.5~$GeV$^2$, which coincides with the value $Q_{\rm min}^2$ that defines
the kinematic cut $Q^2 \ge Q_{\rm min}^2$ for the
data points that are used as input to the fits.
The charm PDF is then generated perturbatively from quarks and gluons by
means of DGLAP evolution, exploiting recent developments in {\tt
  APFEL} which allow the setting of heavy-quark thresholds $\mu_h$
differently from the heavy quark masses $m_h$, such that
$\mu_c=Q_0 > m_c$. Hence a high threshold can be used without having to parametrise the charm PDF \cite{current:work}.
% as done in~\cite{Ball:2016neh}.

The expression for the $\chi^2$  function used for the fits is that
of Ref.~\cite{Aaron:2012qi}, which
includes corrections for possible 
biases from statistical fluctuations and treats the systematic uncertainties
multiplicatively.
Alternative forms that do not include these corrections, such as those
defined in~\cite{Aaron:2009aa,Abramowicz:2015mha},
have also been studied but no significant differences
in the results have been observed.

In this analysis, the parametrised PDFs are the valence distributions
$xu_{v}(x,Q_0^2)$ and $xd_{v}(x,Q_0^2)$, the gluon distribution $xg(x,Q_0^2)$, and the
\textit{u}-type and \textit{d}-type sea-quark distributions,
$x\bar{U}(x,Q_0^2)$, $x\bar{D}(x,Q_0^2)$, where $x\bar{U}(x,Q_0^2) = x\bar{u}(x,Q_0^2)$ and
$x\bar{D}(x,Q_0^2) = x\bar{d}(x,Q_0^2) + x\bar{s}(x,Q_0^2)$.
In addition, the photon distribution $x\gamma(x,Q_0^2)$ is also parametrised at the starting scale.
The following general functional form is adopted:
\begin{equation}
  \label{eq:parametrization}
xf(x, Q_0^2) = Ax^{B}(1-x)^{C}(1+Dx+Ex^{2}) \, ,
\end{equation}
where some of the normalisation parameters, in particular $A_{u_{v}}$,
$A_{d_{v}}$ and $A_{g}$, are constrained by the valence and momentum
sum rules (note that the photon PDF also enters the momentum sum rule).
The parameters $B_{\bar{U}}$ and $B_{\bar{D}}$ are set equal to each
other, so that the two quark sea distributions share a common
small-$x$ behaviour.
Since the measurements used here are not sensitive to the strangeness
content of the proton, strangeness is fixed such that $x\bar{s} (x, Q_0^2) = r_sx\bar{d}(x,Q_0^2)$, where
$r_s=1.0$ is consistent with the ATLAS analysis of inclusive $W$
and $Z$ production~\cite{Aad:2012sb,Aaboud:2016btc}.
The further constraint $A_{\bar{U}} = 0.5 A_{\bar{D}}$ is imposed,
such that $x\bar{u}(x,Q_0^2) \to x\bar{d}(x,Q_0^2)$ as $x \to 0$.  

The explicit form of PDF parametrisation Eq.~(\ref{eq:parametrization})
at the scale $Q_0^2$ is determined by the
technique of saturation of the $\chi^{2}$, namely the number of parameters is increased 
one by one  until the $\chi^{2}$ does not improve
further, employing Wilks' theorem~\cite{Wilks:1938dza}.
Following this method, the optimal parametrisation for the quark and
gluon PDFs found for this analysis is:
\begin{eqnarray}
  \nonumber
  xu_v(x) &&= A_{u_v}x^{B_{u_v}}(1-x)^{C_{u_v}}(1+E_{u_v}x^{2})\, , \\
  \nonumber
xd_v(x) &&= A_{d_v}x^{B_{d_v}}(1-x)^{C_{d_v}}\, , \\
x\bar{U}(x) &&= A_{\bar{U}}x^{B_{\bar{U}}}(1-x)^{C_{\bar{U}}}\, , \\
\nonumber
x\bar{D}(x) &&= A_{\bar{D}}x^{B_{\bar{D}}}(1-x)^{C_{\bar{D}}}\, , \\
\nonumber
\label{eq:param}
xg(x) &&= A_{g}x^{B_{g}}(1-x)^{C_{g}}(1+E_{g}x^{2})\, ,
\end{eqnarray}
while for the photon PDF it is used:
\begin{equation}
x\gamma(x) = A_{\gamma}x^{B_{\gamma}}(1-x)^{C_{\gamma}}(1+D_{\gamma}x+E_{\gamma}x^{2}) \, .
\end{equation}

The parametrisation of the quark and gluon PDFs in
Eq.~(\ref{eq:param}) differs from the one used in the HERAPDF2.0
analysis in various ways.
First of all, a higher value of the input evolution scale
$Q_0^2$ is used, which is helpful to stabilise the fit of the photon PDF.
Second, an additional negative term in the parametrisation of the
gluon is not required here, because of the increased value of $Q_0^2$
which assures the positiveness of the gluon distribution.
Third, the results of the parametrisation scan are different because of the
inclusion of the ATLAS high-mass Drell-Yan cross-section data.

PDF uncertainties are estimated using the Monte Carlo replica
method~\cite{DelDebbio:2004xtd,DelDebbio:2007ee,Ball:2008by},
cross-checked with
the Hessian method~\cite{Pumplin:2001ct} using $\Delta\chi^2=1$.
The former is expected to be more robust than the latter, due to the
potential non-Gaussian nature of the photon PDF
uncertainties~\cite{Ball:2013hta}.
In Section \ref{sec:results} it is shown that these two methods to estimate the PDF uncertainties
on the photon PDF lead to similar results.

In addition, a number of cross-checks have been performed to assess the
impact of various model and parametrisation uncertainties.
For the model uncertainties, variations of the charm mass
between $m_c=1.41$ GeV to 1.53 GeV, of the bottom mass between
$m_c=4.25$ GeV to 4.75 GeV, of the strong coupling constant
$\alpha_s(m_Z)$ between 0.116 to 0.120 are considered, and additionally the
strangeness fraction is decreased down to $r_s=0.75$.
For the parametrisation uncertainties, the impact
of increasing the input parametrisation scale up to $Q_0^2=10$~GeV$^2$ is considered as well as the impact of including additional parameters in Eq.~(\ref{eq:param}).
These extra parameters
make little difference to the $\chi^2$ of the fit, but they can
change the shape of the PDFs in a non-negligible way.
Such additional parameters are  $D_{u_v}$, $D_{\bar{u}}$, $E_{\bar{d}}$, as well
and the extra negative term  in the gluon PDF used in HERAPDF2.0.
The impact of these model and parametrisation uncertainties on the baseline results
is quantified in Sect.~\ref{sec:crosschecks}.

\section{Results}
\label{sec:results}

In this section the
determination of the PDFs from a fit to 
HERA inclusive structure functions and ATLAS high-mass Drell-Yan cross sections,
with an emphasis on the photon PDF is presented.
First the fit quality is assessed and the fit results are compared 
with the experimental data.
Then, the resulting photon PDF is shown and compared with other
recent determinations.
The impact of the high-mass DY data on
the quark PDFs is also quantified.
Following this,
the robustness of the fits of $x\gamma(x,Q^2)$
with respect to varying the model, parametrisation, and procedural
inputs is assessed.
Finally, perturbative stability is addressed by comparing NLO and
NNLO results.

\subsection{Fit quality and comparison between data and fit results}

In the following, the results that will be shown
correspond to those obtained from fitting the
double-differential $\lp m_{ll},y_{ll}\rp$ cross-section distributions.
It has been verified that comparable results are obtained if the
$\lp m_{ll},\Delta\eta_{ll}\rp$ cross-section distributions are fitted instead.

For the baseline NNLO fit, the value $\chi^2_{min}/N_{dof} =
1284./1083$ is obtained where $N_{dof}$ is the number of degrees of
freedom in the fit which is equal to total number of data points minus
number of free parameters.
The contribution from the HERA inclusive  data is 
$\chi^{2}/N_{\rm dat} = 1236/1056$ and
from the ATLAS high-mass DY data is $\chi^2/N_{\rm dat} = 48/48$,
where  $N_{\rm dat}$ the number of the data points for the corresponding
data sample.
These values for  $\chi^{2}/N_{\rm dat}$, together
with the corresponding values for the various
invariant mass $m_{ll}$ bins of the ATLAS  data,
are summarised in
Table~\ref{tab:chi2fit}.
The quality of the agreement with the HERA cross sections is of comparable quality to that found in the HERAPDF2.0 analysis.
Note that in the calculation of the total $\chi^2$ for the
ATLAS data, the correlations between the
different $m_{ll}$ bins have been taken into account.

%%%%%%%%%%%%%%%%%%%%%%%%%%%%%%%%%%%%%%%%%%%%
\begin{table}[t]
  \centering
  \begin{tabular}{|c|c|}
    \hline
    Dataset  &   $\chi^2$ /$N_{\rm dat}$ \\
    \hline
    \hline
    HERA I+II & 1236/1056\\
    \hline
    high-mass DY  116 GeV $\le m_{ll} \le $ 150 GeV  &  9/12 \\
    high-mass DY  150 GeV $\le m_{ll} \le $ 200 GeV  &  15/12 \\
    high-mass DY  200 GeV $\le m_{ll} \le $ 300 GeV  &  14/12 \\
    high-mass DY  300 GeV $\le m_{ll} \le $ 500 GeV  &  5/6 \\
    high-mass DY  500 GeV $\le m_{ll} \le $ 1500 GeV &  4/6 \\
    \hline
    Correlated (high-mass DY) $\chi^2$ & 1.17 \\
    Log penalty (high-mass DY) $\chi^2$  & -0.12 \\
    \hline
    Total  (high-mass DY) $\chi^2/N_{\rm dat}$  & 48/48 \\
    \hline
    \hline
    Combined HERA I+II and high-mass DY $\chi^2/N_{\rm dof}$   & 1284/1083 \\
    \hline
    \end{tabular}
  \caption{The $\chi^{2}/N_{\rm dat}$ in the NNLO fits for the
    HERA inclusive structure functions and for the various
    invariant mass $m_{ll}$ bins of the ATLAS high-mass DY data.
    In the latter case, the contribution to the
    $\chi^2$ arising from the correlated and log-penalty terms are indicated,
    as well as the overall $\chi^2/N_{\rm dof}$ is provided,  where $N_{\rm dof}$ is
    the number of degree of freedom in the fit.
\label{tab:chi2fit}
  }
\end{table}
%%%%%%%%%%%%%%%%%%%%%%%%%%%%%%%%%%%

Figs.~\ref{hmDY_2D_1}--\ref{hmDY_2D_3} then show the
comparison between the results of the NNLO fit,
denoted by {\tt xFitter\_epHMDY},
and the ATLAS data
  for the $(m_{ll},|y_{ll}|)$ double-differential Drell-Yan cross-sections
  as function of $|y_{ll}|$, for the five bins in $m_{ll}$ separately.
  The comparisons are shown both
  on an absolute scale  and as ratios to the central value
  of the experimental data.
  The error bars on the data points correspond to the bin-to-bin uncorrelated
  uncertainties, while the bands
  indicate the size of the correlated systematic uncertainties.
  The solid lines indicate the theory calculations obtained using the results
  of the fit.

%%%%%%%%%%%%%%%%%%%%%%%%%%%%%%
\begin{figure}[t]
\centering
\includegraphics[width=7cm]{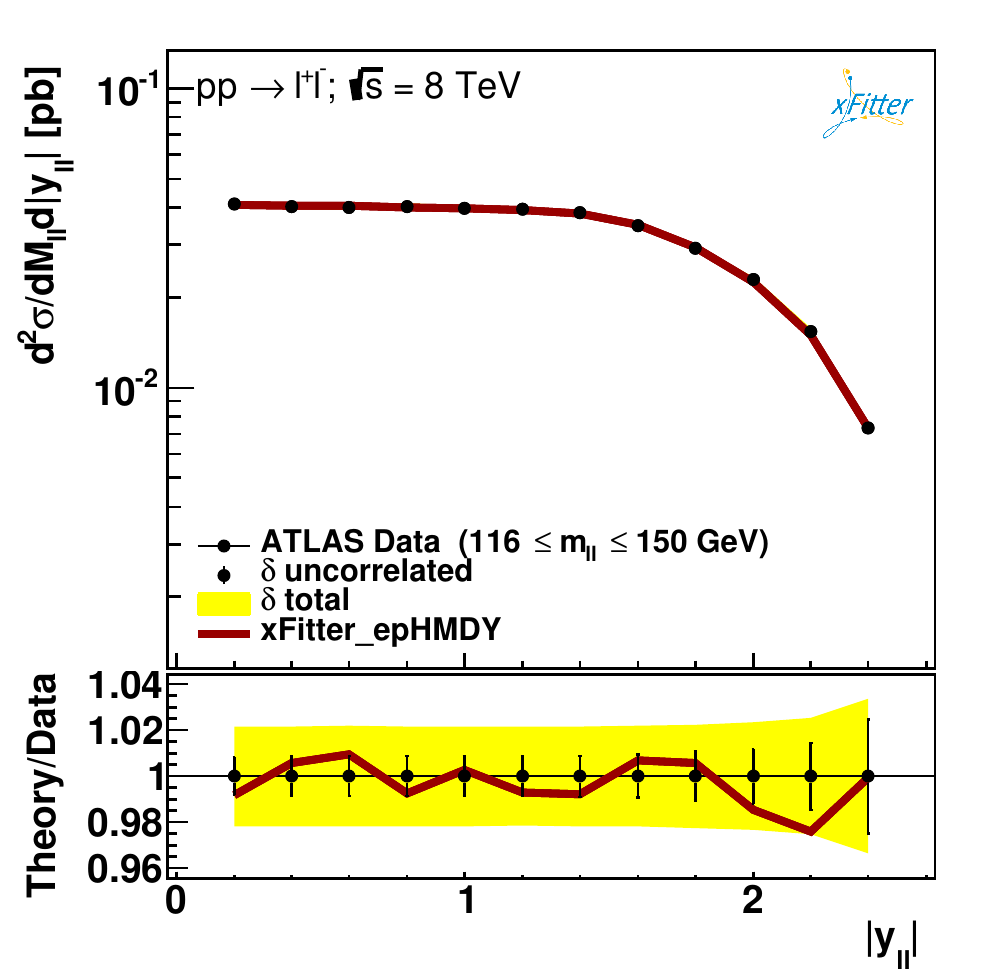}
\includegraphics[width=7cm]{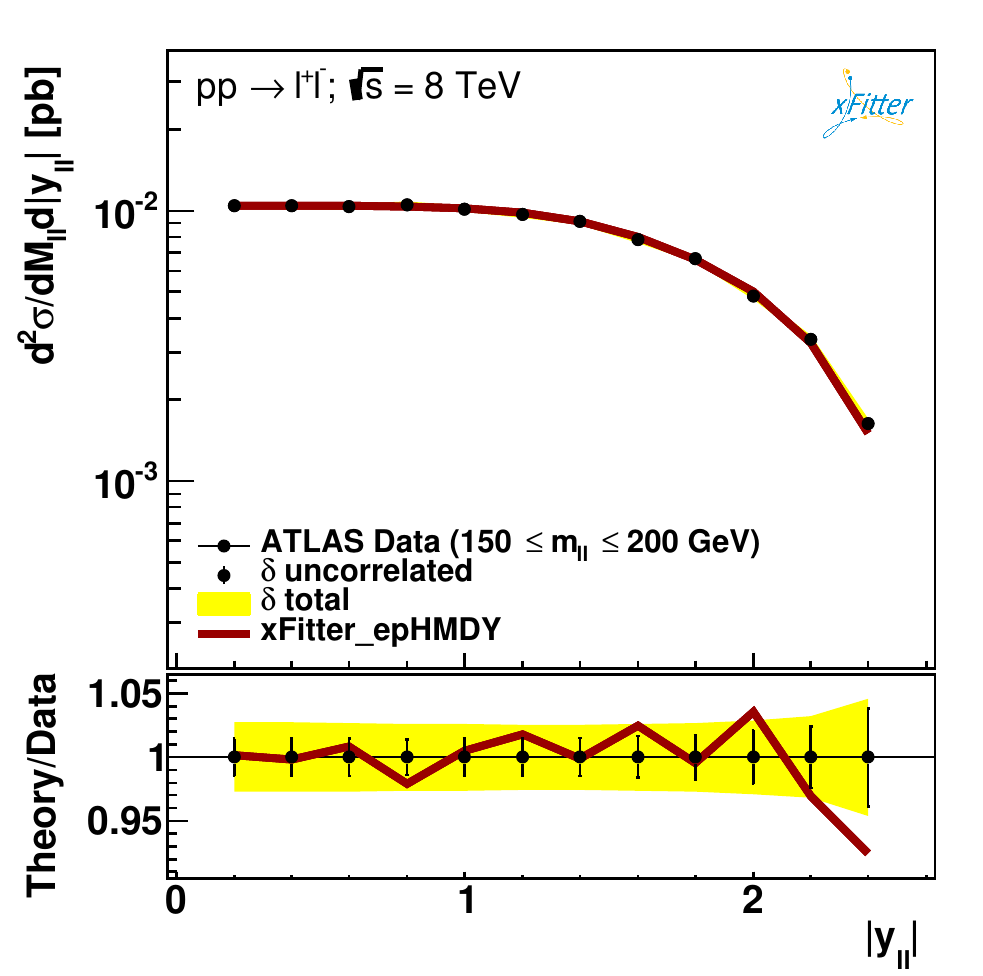}
\caption{Comparison between the results of the fit and the ATLAS data
  for the $(m_{ll},|y_{ll}|)$ double-differential Drell-Yan cross-sections
  as function of $|y_{ll}|$, for the first two $m_{ll}$ bins.
  The comparisons are shown both
  in an absolute scale (upper plots) and as ratios to the central value
  of the experimental data in each $y_{ll}$ bin (lower plots).
  The error bars on the data points correspond to the bin-to-bin uncorrelated
  uncertainties, while the yellow bands
  indicate the size of the correlated uncertainties.
  The solid lines indicate the theory calculations obtained using the results
  of the fit {\tt xFitter\_epHMDY}.
}
\label{hmDY_2D_1}
\end{figure}

%%%%%%%%%%%%%%%%%%%%%%%%%%%%%%
\begin{figure}[t]
\centering
\includegraphics[width=7cm]{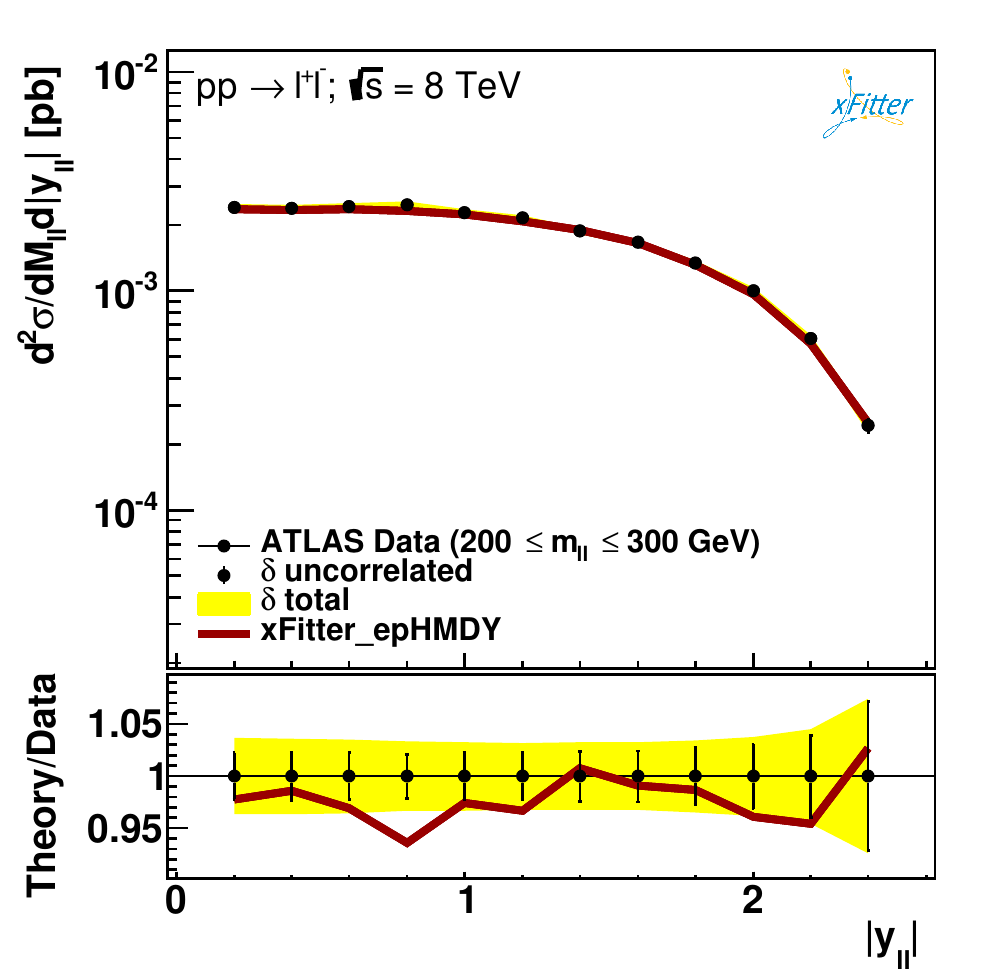}
\includegraphics[width=7cm]{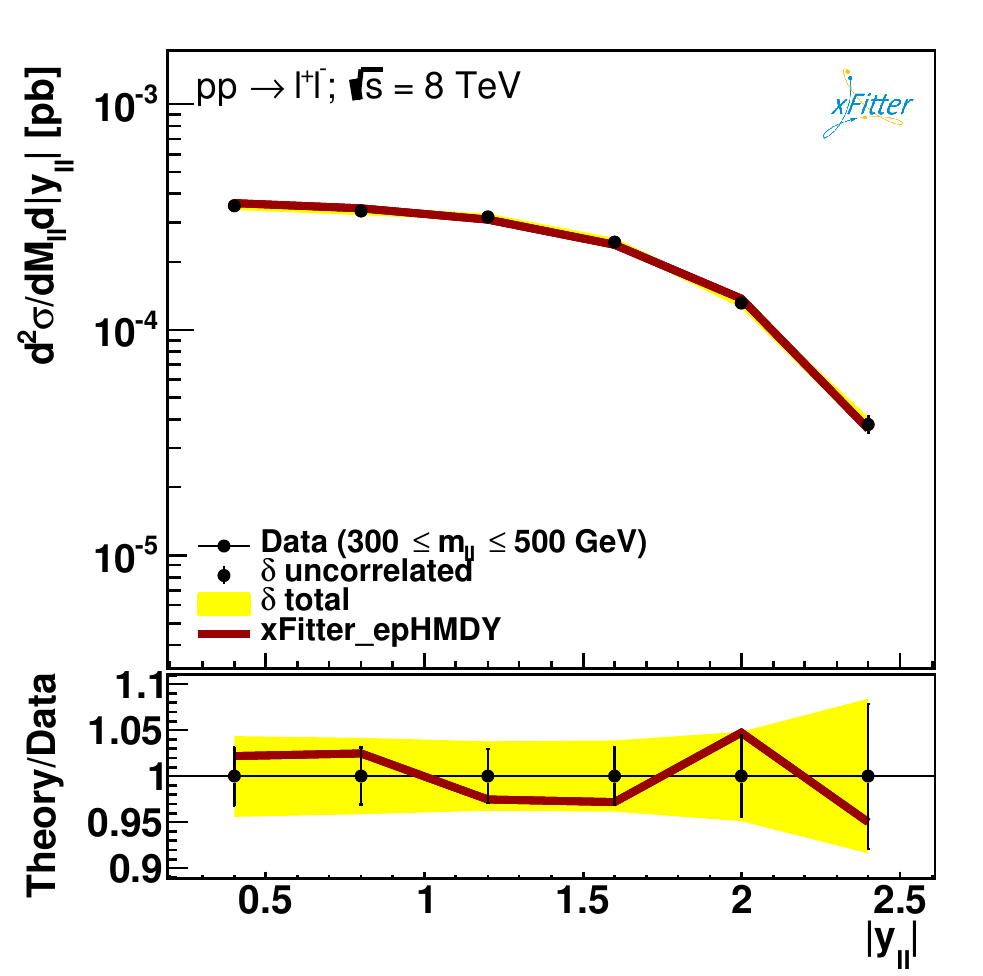}
\caption{Same as Fig.~\ref{hmDY_2D_1} for the third and fourth $m_{ll}$ bins.
}
\label{hmDY_2D_2}
\end{figure}

%%%%%%%%%%%%%%%%%%%%%%%%%%%%%%
\begin{figure}[t]
\centering
\includegraphics[width=7cm]{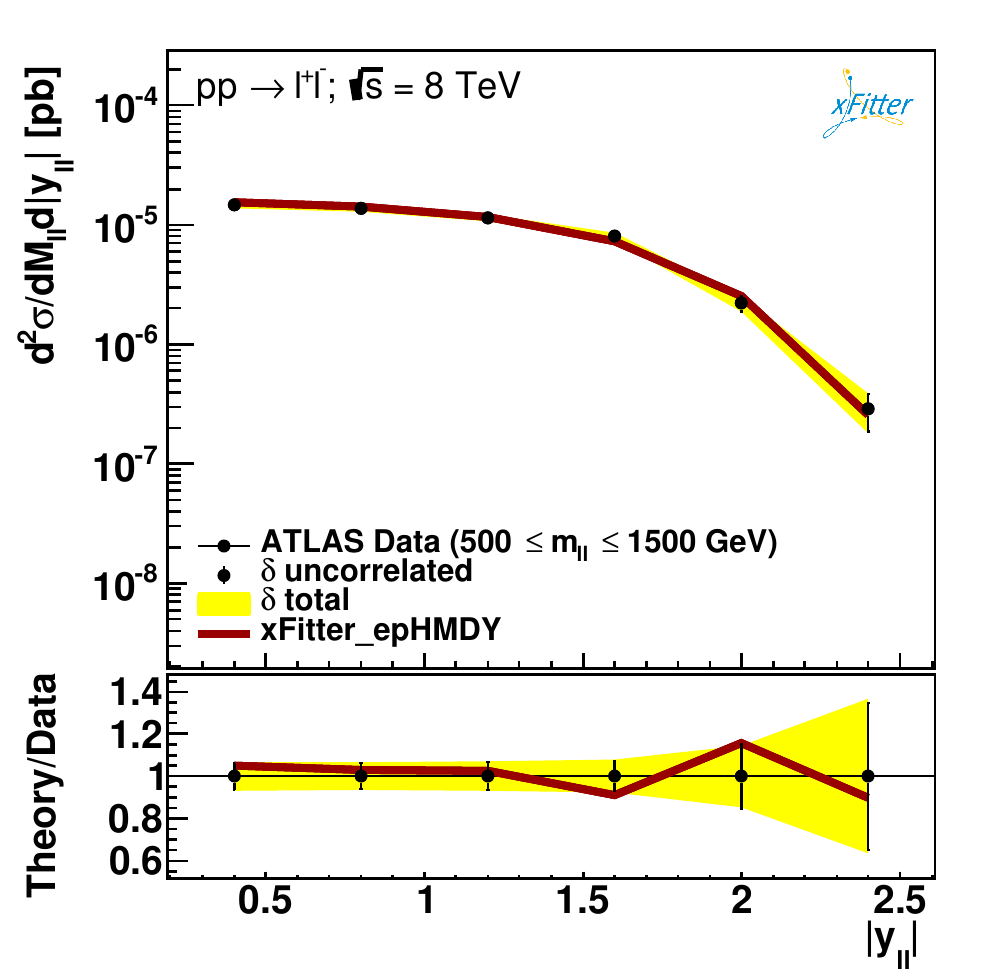}
\caption{Same as Fig.~\ref{hmDY_2D_1} for the highest $m_{ll}$ bin.
}
\label{hmDY_2D_3}
\end{figure}
%%%%%%%%%%%%%%%%%%%%%%%%%%%%%%%

Figs.~\ref{hmDY_2D_1}--\ref{hmDY_2D_3} demonstrate
a good agreement between ATLAS data and the NNLO theory
predictions obtained from the {\tt xFitter\_epHMDY} fit.
This agreement is also quantitatively expressed by the values of the $\chi^2$ reported in
Table~\ref{tab:chi2fit}, where for the ATLAS data a $\chi^2/N_{\rm dat}=1$ is found.
This is particularly remarkable
given the high precision of the data, with total experimental
uncertainties at the few percent level in most of the kinematic range.

\subsection{The photon PDF from LHC high-mass DY data}

%We now present the results of the {\tt xFitter\_epHMDY}
%fit of the photon PDF $x\gamma$ and compare with
%other recent determinations. We will also assess the impact
%of the ATLAS high-mass DY data on the quark and gluon PDFs as compared
%to a fit with only HERA structure functions as input.

%%%%%%%%%%%%%%%%%%%%%%%%%%%%%%%%%%%%%%%%%%%%%%%%%%%%%%%%
\begin{figure}[t]
  \includegraphics[width=7cm]{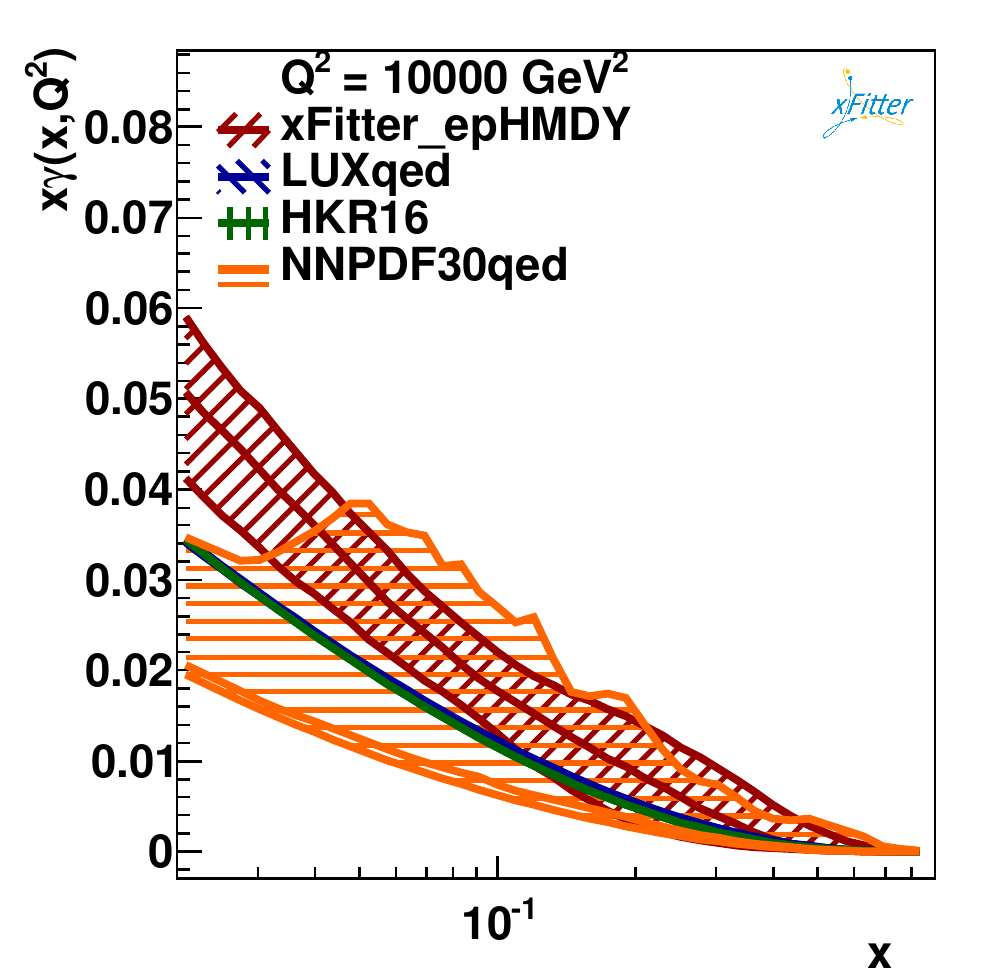}
  \includegraphics[width=7cm]{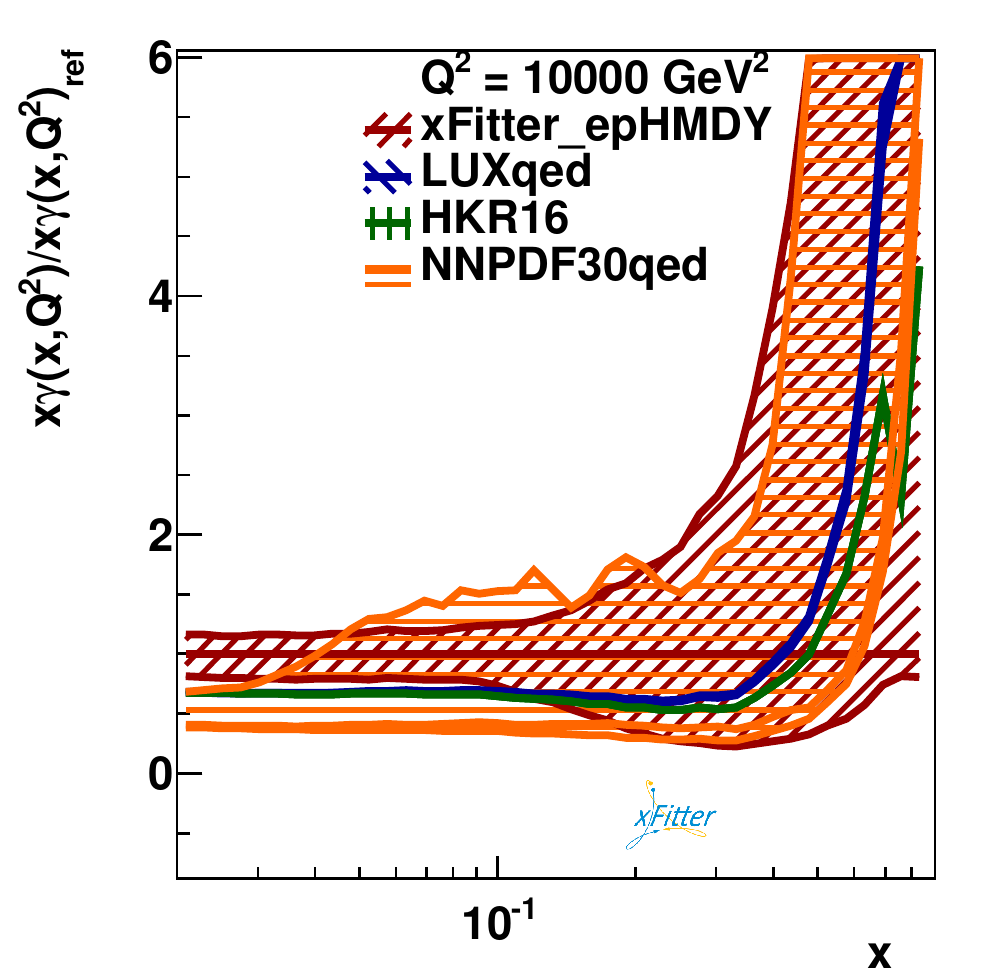} 
  \caption{Left plot: comparison between the photon $x\gamma(x,Q^2)$ at $Q^2=10^4$ GeV$^2$
    from the present NNLO analysis ({\tt xFitter\_epHMDY}) with the
    corresponding results from NNPDF3.0QED, LUXqed and HKR16.
  Right plot: the same comparison, now with the results normalized to the central value
  of {\tt xFitter\_epHMDY}.
  For the present fit, the PDF uncertainties are shown at the 68\% CL obtained from the MC method,
  while model and parametrisation uncertainties are discussed below.
  For HKR16 only the central value is shown, while for LUXqed
  the associated PDF uncertainty band~\cite{Manohar:2016nzj} is included. }
\label{photon_zoom} \label{photon_zoom_ratio}
\end{figure}
%%%%%%%%%%%%%%%%%%%%%%%%%%%%%%%%%%%%%%%%%%%%%%%%%%%%%%%%

In Fig.~\ref{photon_zoom}, the photon PDF, $x\gamma(x,Q^2)$, is shown at
$Q^2=10^4$ GeV$^2$,  and it is compared to the corresponding LUXqed,
HKR16 and NNPDF3.0QED results.
In the left plot the comparison is presented in an absolute scale, while
in the right plot the ratio of
different results normalized to
the central value of the fit is shown.
For the present fit, {\tt xFitter\_epHMDY}, 
the experimental PDF uncertainties at the 68\% confidence level (CL) are obtained from the Monte Carlo method,
 while model and parametrisation uncertainties are discussed below.
Likewise, the  NNPDF3.0QED PDF set is shown the 68\% CL uncertainty band,
while for LUXqed the associated PDF uncertainty band is computed according to the
prescription of Ref.~\cite{Manohar:2016nzj}.
For HKR16, only the central value is available.
The $x$-range in Fig.~\ref{photon_zoom} has been restricted to the region
$0.02 \le x \le 0.9$, since beyond that region there is only limited sensitivity to $x\gamma(x,Q^2)$.

Fig.~\ref{photon_zoom} shows that for $x\ge 0.1$ the four determinations of
the photon PDF are consistent within PDF uncertainties.
For smaller values of $x$, the photon PDF from LUXqed and HKR16 is somewhat smaller than {\tt xFitter\_epHMDY},
but still in agreement at the 2-$\sigma$ level.
This agreement is further improved if the PDF uncertainties in
{\tt xFitter\_epHMDY}
arising from variations of the input parametrisation are added to experimental
uncertainties, as discussed in Sec.~\ref{sec:crosschecks}.
Moreover, the results of this work and NNPDF3.0QED agree at the 68\% CL for $x\ge 0.03$,
and the agreement extends to smaller values of $x$ once the parametrisation
uncertainties in {\tt xFitter\_epHMDY} are accounted for.
The LUXqed and the HKR16 calculations of $x\gamma(x,Q^2)$ are very close
to each other across the entire range of $x$.

Fig.~\ref{photon_zoom} shows that
for $0.04 \le x \le 0.2$ the present analysis  exhibits smaller PDF
uncertainties as compared to those from  NNPDF3.0QED.
Indeed, the experimental uncertainty on the {\tt xFitter\_epHMDY}
turns out to be at the  $\sim 30\%$ level for $x\le 0.1$.
At larger $x$ it increases rapidly
specially in the positive direction.
The reason for this behaviour at large $x$ can be understood by recalling that
variations of $x\gamma(x,Q^2)$ in the negative
direction are constrained by positiveness.
The limited sensitivity of the ATLAS data  does not allow a determination of $x\gamma(x,Q^2)$ with uncertainties
competitive with those of LUXqed, which are at the few percent level.

It is also interesting to assess the impact of the high-mass Drell-Yan 8 TeV measurements on
the light quark and gluon PDFs.
For this purpose, the fits have been repeated freezing the photon PDF to the
{\tt xFitter\_epHMDY} shape. 
This is necessary because
HERA inclusive data alone, which are the 
benchmark for this comparison, have no sensitivity to the photon
PDF.
This way, 
a meaningful comparison between the quark and gluon PDFs
from a HERA-only
baseline and the HERA+HMDY fit can be performed.
% with the effects of
%the photon PDF being factored out.

This comparison is shown in Fig.~\ref{fig:QCDfit} for the up and down
antiquarks $x\bar{u}(x,Q^2)$ and $x\bar{d}(x,Q^2)$, for which the effect of the high-mass Drell-Yan data is
expected to be most pronounced, since HERA inclusive cross sections
provide little information on quark flavour separation.
In Fig.~\ref{fig:QCDfit}, the $x\bar{u}(x,Q^2)$ and $x\bar{d}(x,Q^2)$ together
with the associated MC uncertainties have been computed
at the initial parametrisation scale of $Q^2=7.5$ GeV$^2$ and
are shown as ratios to the central value of the {\tt xFitter\_epHMDY}
fit.
The modifications in the medium
and large-$x$ antiquark distributions from the high mass DY data are
rather moderate.
%
%For $x\bar{d}$, the impact of the ATLAS DY data is negligible, while
%for $x\bar{u}$ it is somewhat larger, though still quite small,
%for $x\ge 0.05$.
%
It has been verified that the same conclusions can be derived
from fits obtained by switching
off the QED effects for both the HERA only fits and the HERA+HMDY fits.
Therefore, while the ATLAS high-mass Drell-Yan measurements have a significant constraint
on the photon PDF, their impact on the quark and gluon PDFs is moderate.

%%%%%%%%%%%%%%%%%%%%%%%%%%%%%%
\begin{figure}[t]
\centering
\includegraphics[width=7cm]{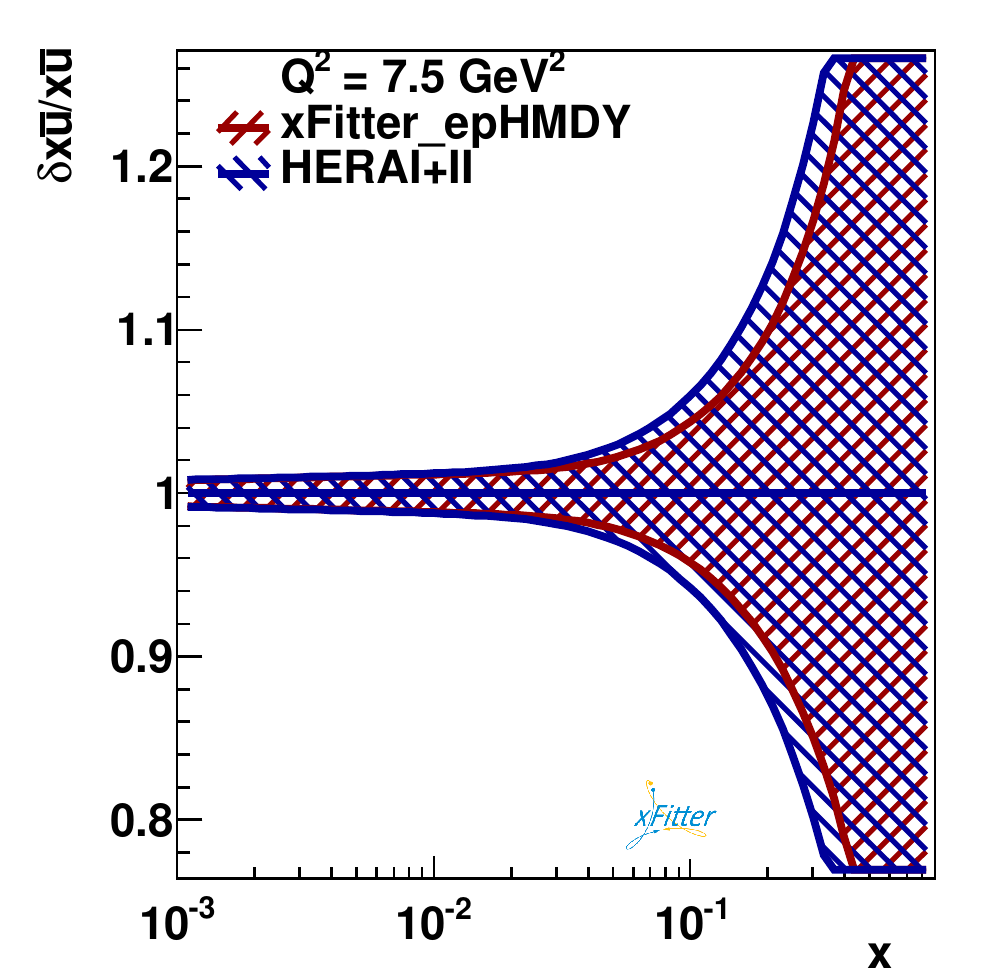}
\includegraphics[width=7cm]{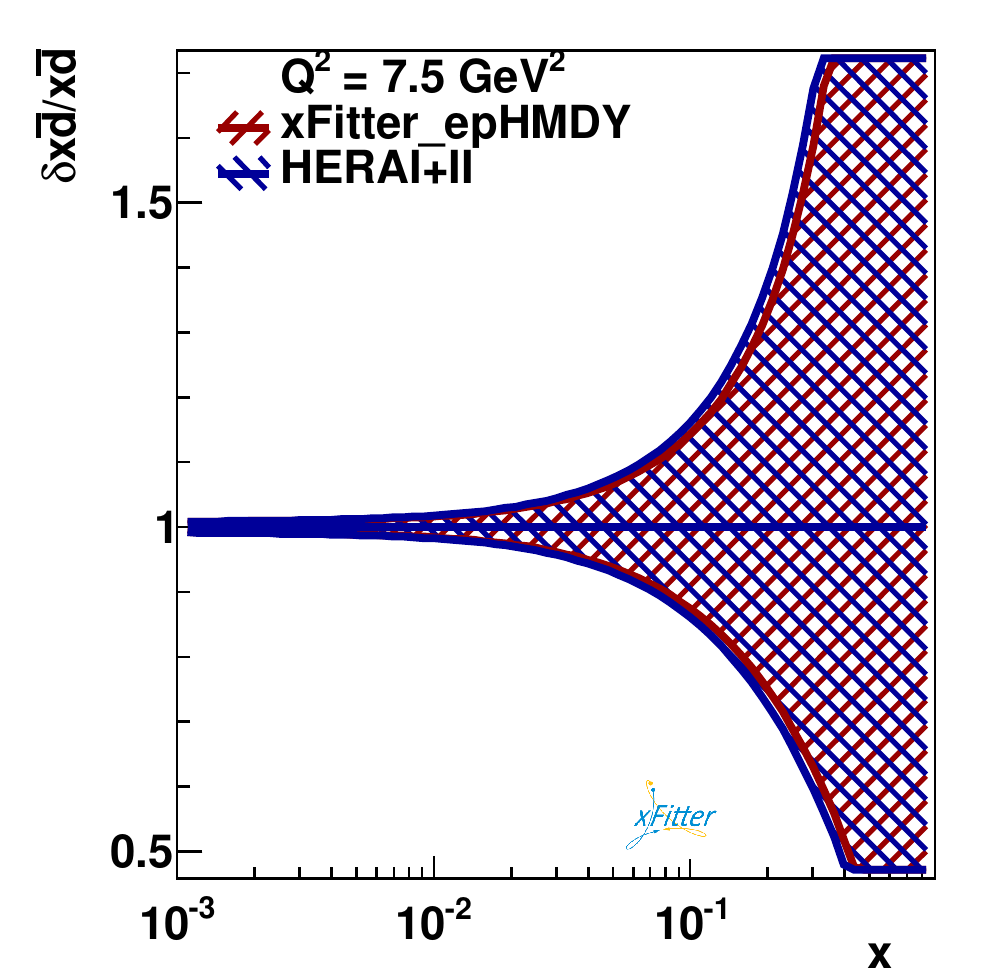} 
\caption{The impact of the ATLAS high-mass 8 TeV Drell-Yan measurements
  on the $x\bar{u}$ and $x\bar{d}$ sea quark PDFs at the input
  parametrisation scale $Q^2=7.5$ GeV$^2$.
  The results are shown normalized to the central value of {\tt xFitter\_epHMDY}. 
  }
\label{fig:QCDfit}
\end{figure}
%%%%%%%%%%%%%%%%%%%%%%%%%%%%%%%

\subsection{Robustness and perturbative stability checks}
\label{sec:crosschecks}

Following the presentation of the main result of this work, the {\tt
  xFitter\_epHMDY} determination of the
photon PDF $x\gamma(x,Q^2)$, the robustness of this determination
with respect to a number of variations is assessed.
Firstly, variations in the values of the input
physical parameters, such as $\alpha_s$ or the charm mass are explored.
Secondly, variations of the choices made for the PDF input parametrisation
are considered.
Finally,
variations associated to different methodological choices in
the fitting procedure are quantified.
In each case, one variation at a time is performed 
and compared with the central value
of $x\gamma(x,Q^2)$ and its experimental PDF
uncertainties computed using the Monte Carlo method.

%%%%%%%%%%%%%%%%%%%%%%%%%%%%%%                                          
\begin{figure}[t]
\centering
\includegraphics[width=7cm]{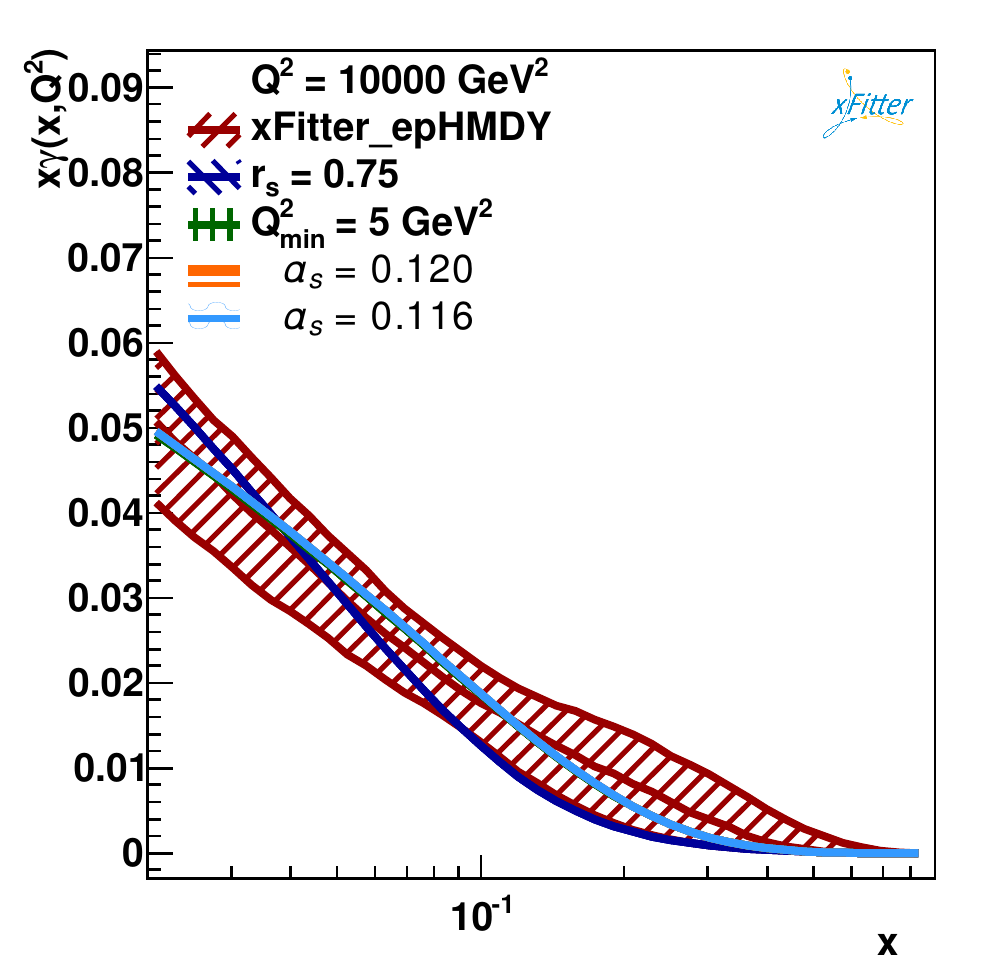}
\includegraphics[width=7cm]{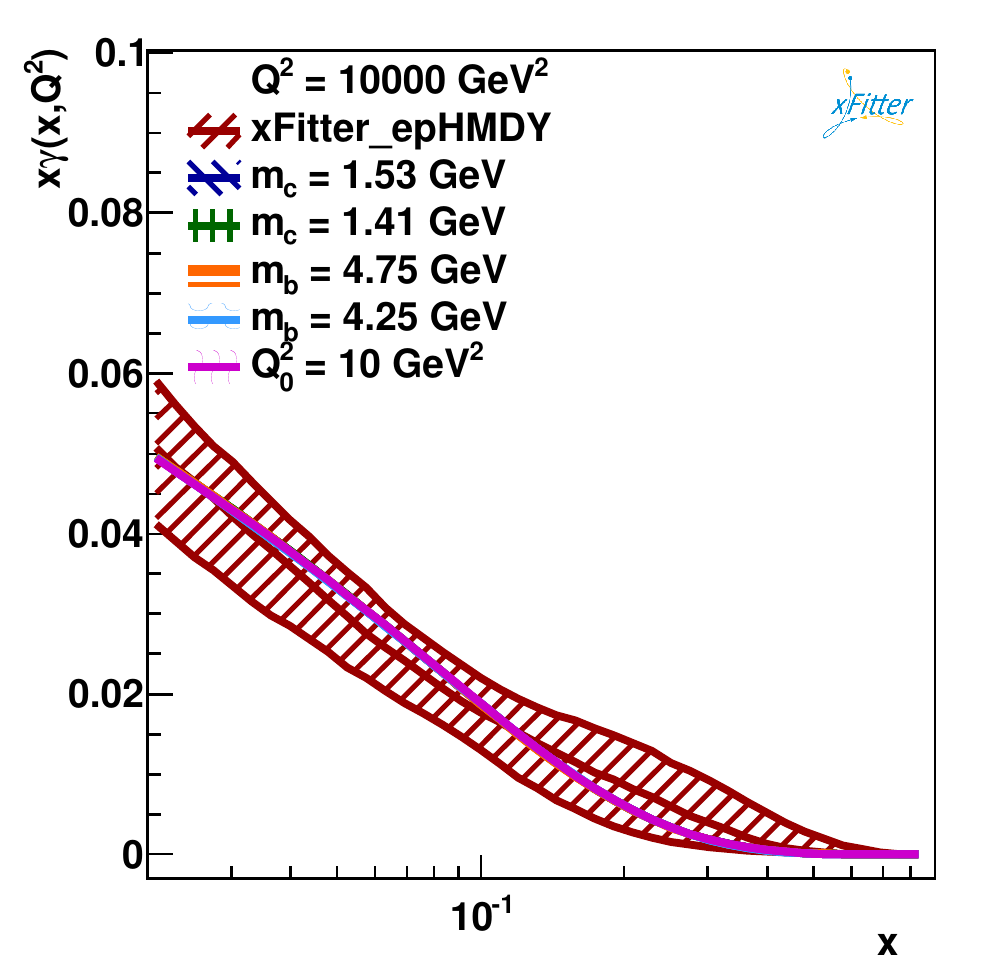}
\caption{Comparison between the baseline determination of
  $x\gamma(x,Q^2)$ at $Q^2=10^4$ GeV$^2$ in the present
  analysis, {\tt xFitter\_epHMDY},
  with the central value of a number of fits for which one input parameter has been varied.
  The following variations have been considered: $r_s=0.75$, $Q^2_{\rm min}=5$ GeV$^2$, $\alpha_s=0.116$ and
  0.118 (left plot); and
  $m_c=1.41$ and 1.53 GeV, $m_b=4.25$ and 4.75 GeV, and $Q_0^2=10$ GeV$^2$ (right plot).
  See text for more details about these variations.
}
\label{fig:model}
\end{figure}
%%%%%%%%%%%%%%%%%%%%%%%%%%%%%%

First the impact of uncertainties associated to either
the choice of input physical parameters or of specific settings
adopted in the fit is considered.
Fig.~\ref{fig:model} shows the comparison between the
{\tt xFitter\_epHMDY} 
determination of $x\gamma(x,Q^2)$ at $Q^2=10^4$ GeV$^2$, including the experimental MC
uncertainties, with the central
value of those fits for which a number of variations have been
performed.
Specifically:
\begin{itemize}
\item The strong coupling constant is varied by $\delta \alpha_s=\pm 0.002$ around the central value.
\item The ratio of strange to non-strange light quark PDFs is decreased to $r_s=0.75$ instead of $r_s=1$.
\item The value of the charm mass is varied between $m_c=1.41$ GeV and $m_c=1.53$ GeV,
  and that of the bottom mass between $m_b=4.25$ GeV and $m_b=4.75$ GeV.
\item The minimum value $Q_{\rm min}^2$ of the fitted data is decreased down to $5$ GeV$^2$.
\item The input parametrisation scale $Q_0^2$ is raised to $10$ GeV$^2$ as compared
  to the baseline value of $Q_0^2=7.5~$GeV$^2$.
\end{itemize}
The results of Fig.~\ref{fig:model} highlight that in all cases effect
of the variations considered here is contained within (and typically much smaller than) 
the experimental PDF uncertainty bands of the reference fit.
The largest variation comes from the strangeness ratio $r_s$, where the resulting
central value turns out to be at the bottom end of the PDF uncertainty band for $x\ge 0.1$.

Another important check of the robustness of the present determination of
$x\gamma(x, Q^2)$ can be obtained by comparing the baseline fit with further
fits where a number of new free parameters are allowed in the PDF
parametrisation, in addition to those listed in Eq.~(\ref{eq:param}).
Fig.~\ref{fig:param} shows the impact of three representative
variations (others have been explored, leading to smaller
differences): more flexibility to the gluon distribution, allowing it
to become negative at the initial scale (labeled by ``${\rm neg}$''), in addition to $D_{u_v}$,
and then $D_{\bar{u}}+D_{\bar{d}}$.
As before, all variations are contained within the experimental PDF uncertainty
bands, though the impact of the parametrisation variations is typically larger
than that of the model variations: in the case of the
${\rm neg}+D_{\bar{u}}+D_{\bar{d}}$ variations, the central value is
at the lower edge of the PDF uncertainty band in the entire range
of $x$ shown.
%
%Importantly, once this additional source of PDF uncertainty arising from the
%parametrisation variations is accounted for, the agreement of the {\tt xFitter\_epHMDY}
%fit 
%with the LUXqed and HKR16  determinations
%shown in Fig.~\ref{photon_zoom} improves in the
%region around $x\simeq 0.02$.

A cross-check of the robustness of the estimated
experimental uncertainty of the photon PDF in this analysis is provided by the
comparison of the Monte Carlo and Hessian methods.
Fig.~\ref{fig:photon_mc_vs_hessian} shows this comparison
indicating a reasonable agreement between the two methods.
In particular, the central values of the photon obtained with the two
fitting techniques are quite similar to each other.
As expected, the MC uncertainties tend to be larger than the Hessian ones,
specially in the region $x\gsim 0.2$, indicating deviations with respect
to the Gaussian behaviour of the photon PDF.

%%%%%%%%%%%%%%%%%%%%%%%%%%%%%%%%%%%
\begin{figure}[t]
\centering
\includegraphics[width=7cm]{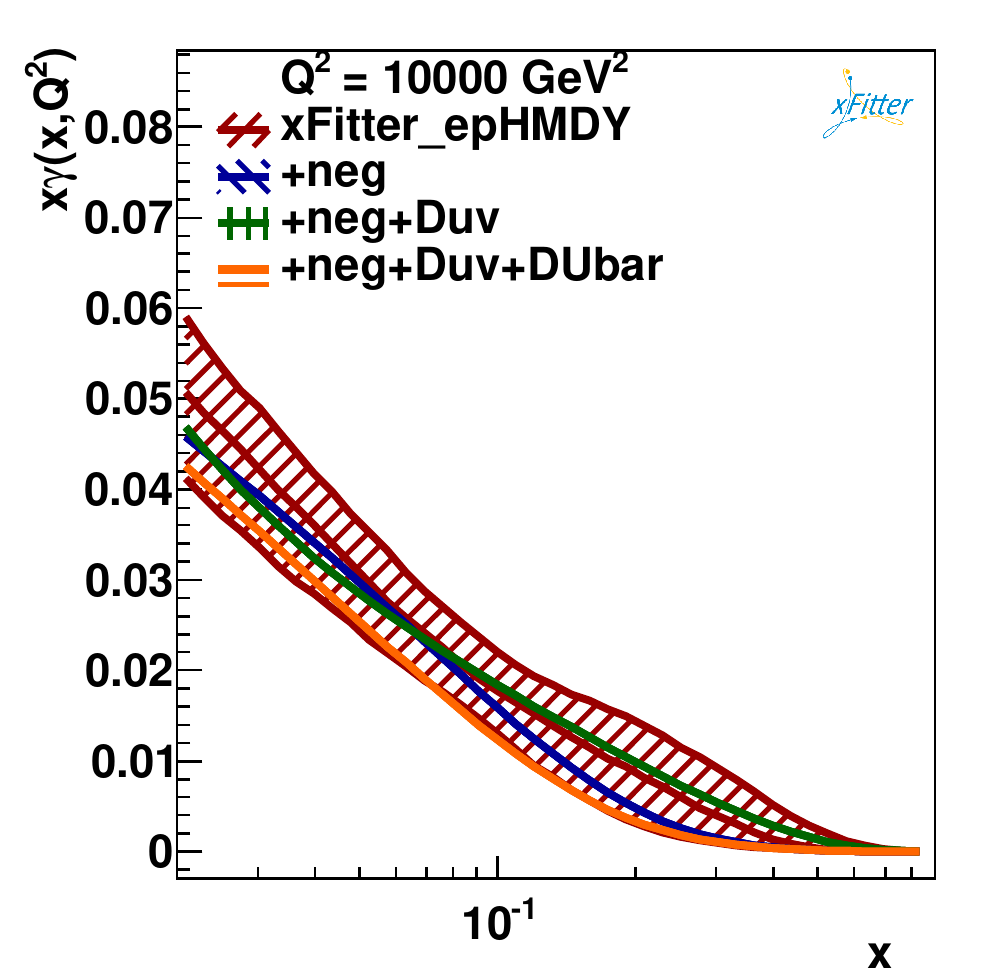}
\includegraphics[width=7cm]{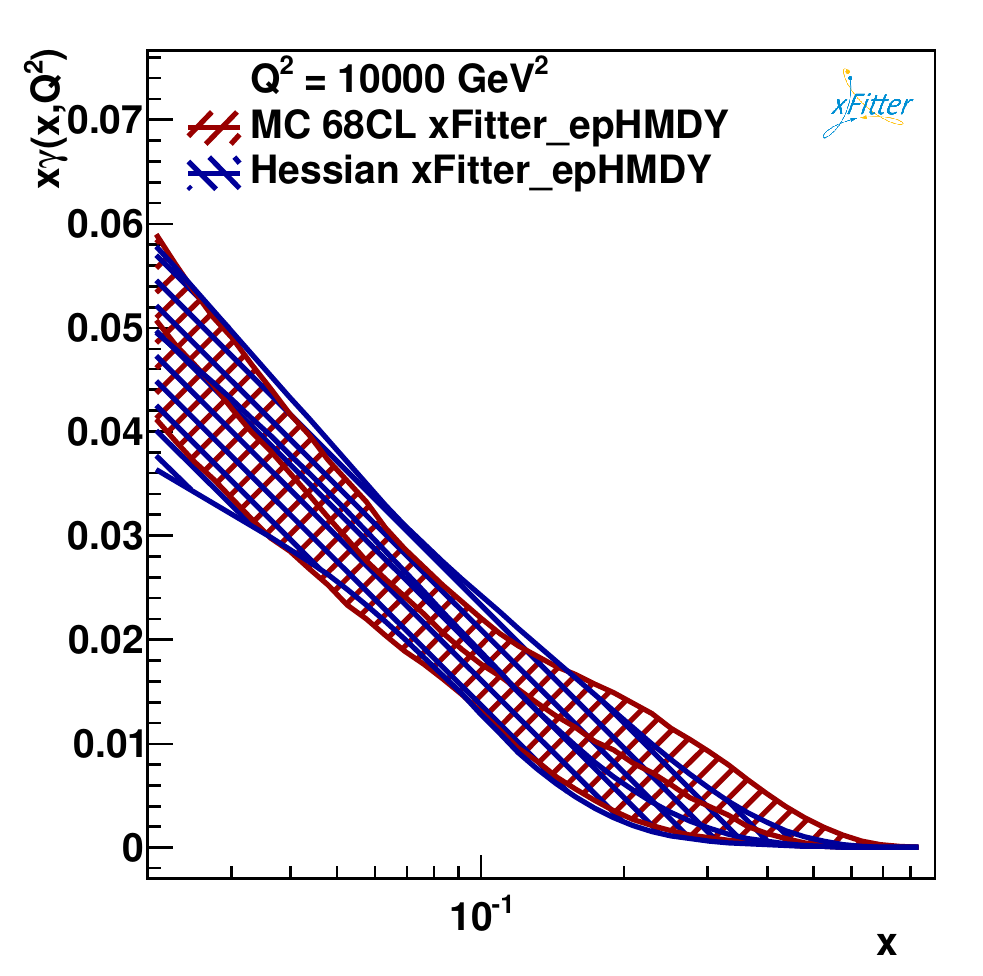} 
\caption{Left: the impact on the photon PDF $x\gamma(x,Q^2)$
  from {\tt xFitter\_epHMDY}
  in fits where a number of additional free parameters are allowed
  in the PDF parametrisation Eq.~(\ref{eq:param}).
  The parametrisation variations that have been explored
  are: more flexibility to the gluon distribution, allowing
  it to become negative
  (labeled by ``${\rm neg}$''), adding on top $D_{u_v}$, and then
  adding $D_{\bar{u}}+D_{\bar{d}}$.
 Right: comparison between the {\tt xFitter\_epHMDY} determinations obtained with the
 Monte Carlo (baseline) and with the Hessian methods, where in
  both cases the PDF error band  shown corresponds to the 68\% CL uncertainties.  }
\label{fig:param}
\label{fig:photon_mc_vs_hessian}
\end{figure}
%%%%%%%%%%%%%%%%%%%%%%%%%%

To complete these studies, an interesting exercise is to quantify the perturbative stability of
the {\tt xFitter\_epHMDY}
determination of the photon PDF $x\gamma(x,Q^2)$ with respect to the inclusion
of NNLO QCD corrections in the analysis.
To study this, Fig.~\ref{fig:nlo_vs_nnlo} shows a
comparison between the baseline fit of $x\gamma(x,Q^2)$, based on NNLO
QCD and NLO QED theoretical calculations, with the central value resulting from a
corresponding fit
based instead on NLO QCD and QED theory.
In other words, the QED part of the calculations is identical in both cases.
For the NNLO fit, only the experimental PDF uncertainties, estimated
using the Monte Carlo method, are shown.
From the comparison of Fig.~\ref{fig:nlo_vs_nnlo}, it is clear that the
fit of $x\gamma(x,Q^2)$ exhibits a reasonable perturbative stability,
since the central value of the NLO fit is always contained in the
one-sigma PDF uncertainty band of the baseline {\tt xFitter\_epHMDY} fit.
The agreement between the two fits is particularly good for
$x\gsim 0.1$, where the two central values are very close to each
other.
This comparison is shown at low scale, $Q^2=7.5$ GeV$^2$ and high scales $Q^2=10^4$ GeV$^2$,
indicating that perturbative stability is not scale dependent.

%%%%%%%%%%%%%%%%%%%%%%%%%%%%%%%%%%%%%%%%%%%%%%%%%%%%%%%%%%%%%%%%%%%%%%%%%%%%%%%%%%%%%%%%%%%%%
%%%%%%%%%%%%%%%%%%%%%%%%%%%%%%%%%%%%%%%%%%%%%%%%%%%%%%%%%%%%%%%%%%%%%%%%%%%%%%%%%%%%%%%%%%%%%
\begin{figure}[t]
\centering
\includegraphics[width=7cm]{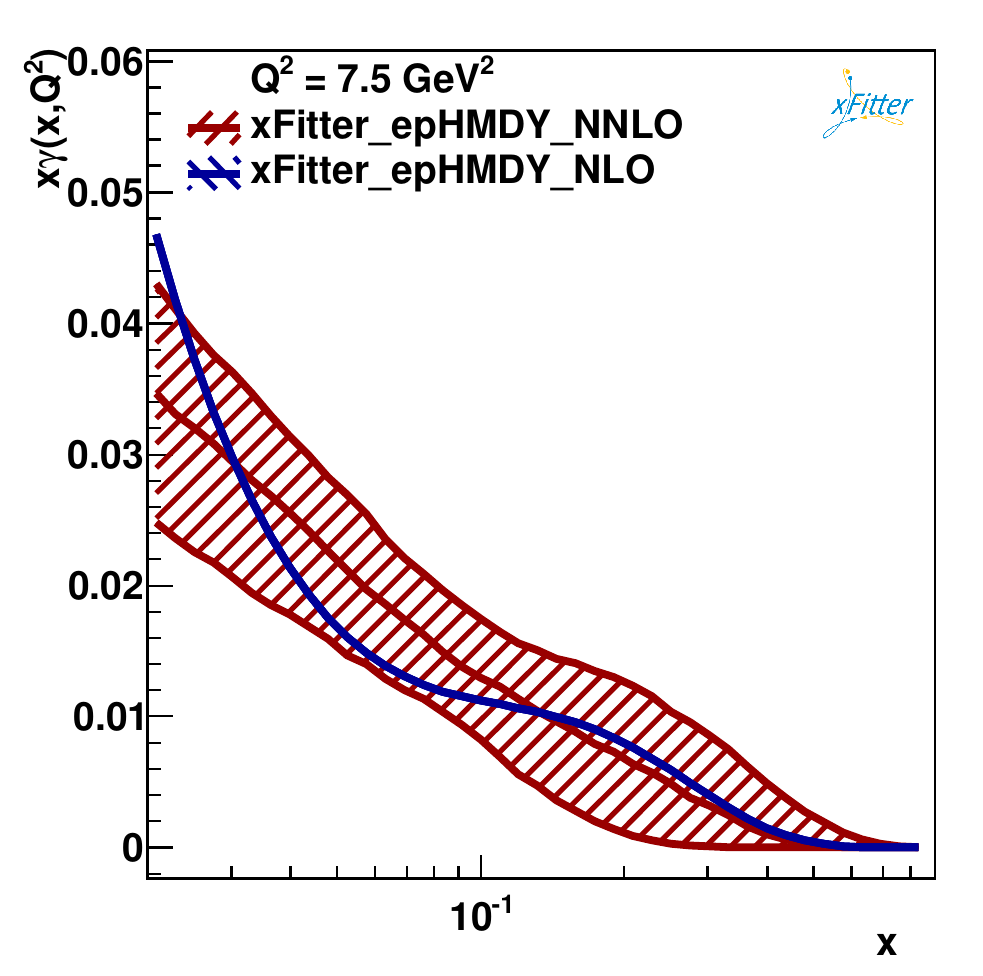}
\includegraphics[width=7cm]{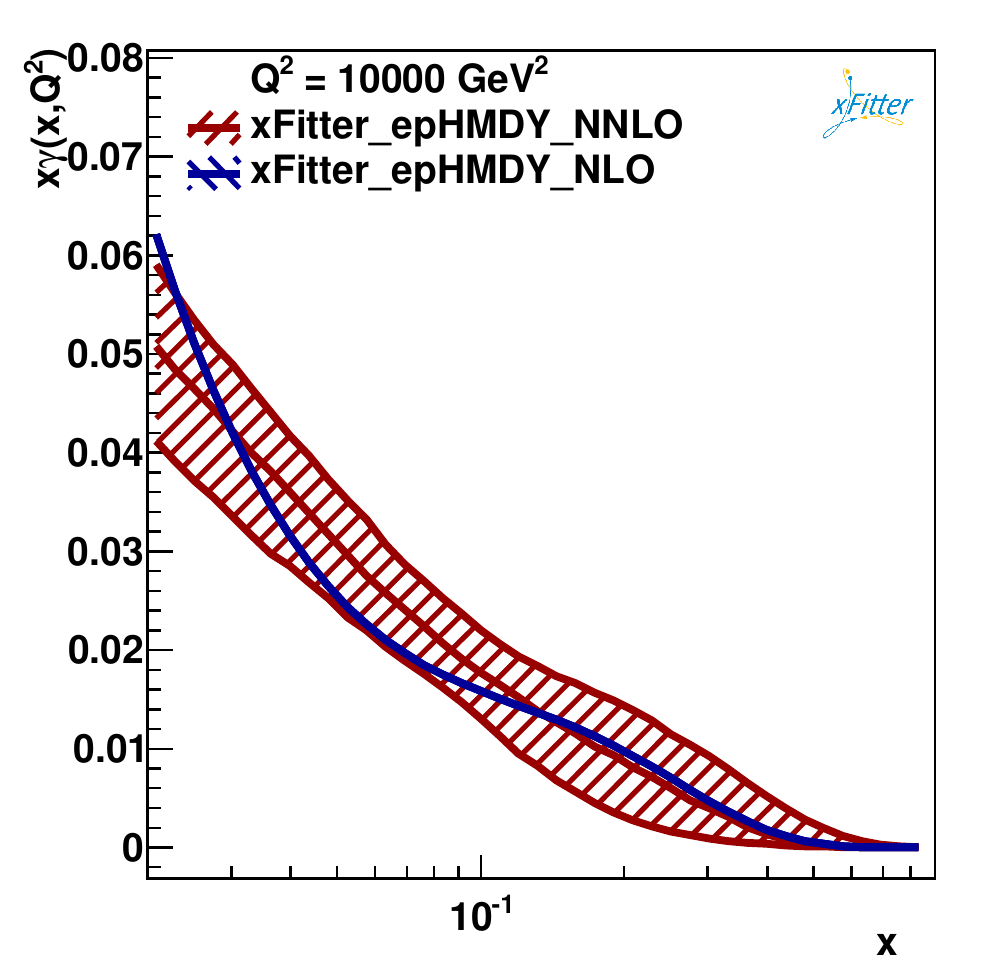}
\caption{Left plot: comparison between the reference {\tt xFitter\_epHMDY}
  fit of $x\gamma(x,Q^2)$, based on  NNLO QCD and NLO QED theoretical
  calculations, with the central value of the corresponding fit based
  on NLO QCD and QED theory, at $Q^2=7.5$ GeV$^2$.
  In the former case, only the experimental Monte Carlo PDF
  uncertainties are shown.
  Right plot: same comparison, now presented at the higher scale of $Q^2=10^4$ GeV$^2$.  }
\label{fig:nlo_vs_nnlo}
\end{figure}
%%%%%%%%%%%%%%%%%%%%%%%%%%%%%%%%%%%%%%%%%%%%%%%%%%%%%%%%%%%%%%%%%%%%%%%%%%%%%%%%%%%%%%%%%%%%
%%%%%%%%%%%%%%%%%%%%%%%%%%%%%%%%%%%%%%%%%%%%%%%%%%%%%%%%%%%%%%%%%%%%%%%%%%%%%%%%%%%%%%%%%%%%

\section{Summary}
\label{sec:conclusions}

In this work, a new determination of the photon PDF from
a fit of HERA inclusive DIS structure functions supplemented
by ATLAS data on high-mass Drell-Yan cross
sections has been presented, based on the {\tt xFitter} framework.
As suggested by a previous reweighting analysis~\cite{Aad:2016zzw}, this
high-mass DY data provides significant constraints on the photon PDF,
allowing a determination of $x\gamma(x, Q^2)$ with uncertainties at the
30\% level for $0.02 \le x \le 0.1$.
The results of the present study, dubbed {\tt xFitter\_epHMDY}, are in agreement
and exhibit smaller PDF uncertainties that the only
other existing photon PDF fit from LHC data, the 
NNPDF3.0QED analysis, based on previous LHC Drell-Yan measurements.

The results are in agreement within uncertainties with two recent
calculations of the photon PDF, LUXqed and HKR16.
% in the region of $x$
%where the ATLAS DY data is expected to have sensitivity.
%
For $x\ge 0.1$, the agreement is at the 1-$\sigma$ level already including
only the experimental MC uncertainties, while for $0.02 \ge x \ge 0.1$ it is
important to account for parametrisation uncertainties.
The findings indicate that a direct determination of the
photon PDF from hadron collider data
is still far from being competitive with the LUXqed and HKR
calculations, which are based instead on precise measurements of the inclusive DIS
structure functions of the proton.
%
%However, the {\tt xFitter\_epHMDY} analysis provides further evidence that the information
%on the photon PDF provided by the most sensitive LHC data available so far
%is fully consistent with these two independent calculations.

The results of this study, which
 are available upon
 request in the {\tt LHAPDF6} format~\cite{Buckley:2014ana},
 have been made possible by a number of
technical developments that should be of direct application for future
PDF fits accounting for QED corrections.
First of all, the full NLO
QED corrections to the DGLAP evolution equations and the DIS
structure functions have been implemented in the  {\tt APFEL} program.
Moreover, our results illustrate the flexibility of the {\tt
  xFitter} framework to extend its capabilities beyond the traditional
quark and gluon PDF fits.
Finally, the extension of {\tt aMCfast} and {\tt APPLgrid} to allow
for the presence of photon-initiated channels in the calculations
provided by {\tt MadGraph5\_aMC@NLO}, significantly streamlines the
inclusion of future LHC measurements in PDF fits with QED corrections by
consistently including diagrams with initial-state photons.
All these technical improvements will certainly be helpful for future
studies of the photon content of the proton.\\

{\bf Acknowledgements}.
We thank L.~Harland-Lang for providing us the {\tt LHAPDF6} grid of
the HKR16 photon determination.
We thank A.~Sapronov for providing support to the {\tt xFitter} platform.
%
%We thank R.~Sadykov for early cross checks of the QED evolution program in {\tt xFitter}.
%
We thank P.~Starovoitov for discussions related to the {\tt APPLgrid} files.
We thank M.~Zinser for discussions related to the ATLAS data.
We thank M.~Dyndal for discussions on the scale choices for the photon induced contributions.
The work of V.~B., F.~G., J.~R. has been supported by the European
Research Council Starting Grant ``PDF4BSM".
The work of S.~C. is supported by the HICCUP ERC Consolidator grant (614577).
A.~L. thanks for the support from the Mobility Plus grant no. 1320/MOB/IV/2015/0.
The work of F.~O. has been supported by the US DoE Grant DE-SC0010129.
%
%V.~R. acknowledges support from the Department of Physics at the University of
%Oxford, where part of this work was completed.
%
The work of P.~S and R.~S. has been supported by the BMBF-JINR cooperation.
We are grateful to the DESY IT department for their support of the {\tt xFitter} developers.

\appendix
\section{Implementation of NLO QED corrections in APFEL}
\label{sec:appendixAPFEL}

In this appendix, the details of the implementation of the
combined NLO QCD+QED corrections in the {\tt APFEL} program are presented.
As discussed in Ref.~\cite{Bertone:2013vaa}, the implementation of the
LO QED corrections to the DGLAP evolution equations includes many
simplifications, in particular the fact that QED and QCD corrections
do not mix and therefore the DGLAP equations, as well as the evolution
equations for the running of the $\alpha_s$ and $\alpha$ couplings,
are decoupled.
When increasing the perturbative accuracy to NLO,
this property does not hold anymore, and QED and QCD
contributions mix both in the DGLAP and in the coupling evolution
equations.
On top of this complication, QED corrections introduce the presence of
diagrams in which a real photon is present either in the initial or in
the final state, and these have to be included in the computation of the
DIS structure functions.

In the following, the discussion starts by considering how to generalize the equations for
the running of the QCD and QED couplings, finding that the mixed
QCD+QED terms have a negligible impact.
Then the extension of the DGLAP evolution equations
to account for the complete NLO QCD+QED effects is discussed.
Finally, the modifications introduced by the NLO QED
corrections in both the neutral-current and the charged-current DIS
structure functions are discussed together with those that lead to the 
appearance of photon-initiated
contributions.

\subsection{Evolution of the couplings}

As mentioned above, the NLO QCD+QED corrections induce the presence of
mixed terms in the evolution equations of $\alpha_s$ and $\alpha$.
In practice, the QCD $\beta$-function receives corrections
proportional to $\alpha$ and, vice-versa, the QED $\beta$-function
receives corrections proportional to $\alpha_s$, in such a way that
the coupling evolution equations read:
\begin{equation}\label{CoupledEq}
\begin{array}{rcl}
\displaystyle \mu^2\frac{\partial \alpha_s}{\partial \mu^2} &=& \displaystyle
                                                \beta^{\rm QCD}(\alpha_s,\alpha)\,,\\
\\
\displaystyle \mu^2\frac{\partial \alpha}{\partial \mu^2} &=& \displaystyle \beta^{\rm QED}(\alpha_s,\alpha)\,.
\end{array}
\end{equation}
As a consequence, these evolution equations form a set of coupled
differential equations. Up to three loops ($i.e.$ NLO), the
$\beta$-functions can be expanded as:
\begin{equation}
\beta^{\rm QCD}(\alpha_s,\alpha) = -\alpha_s\left[\beta_0^{(\alpha_s)}\left(\frac{\alpha_s}{4\pi}\right)+\beta_1^{(\alpha_s\alpha)}\left(\frac{\alpha_s}{4\pi}\right) \left(\frac{\alpha}{4\pi}\right)+\beta_1^{(\alpha_s^2)}\left(\frac{\alpha_s}{4\pi}\right)^2+\dots\right]\,,
\end{equation}
and:
\begin{equation}
\beta^{\rm QED}(\alpha_s,\alpha) = -\alpha\left[\beta_0^{(\alpha)}\left(\frac{\alpha}{4\pi}\right)+\beta_1^{(\alpha\alpha_s)}\left(\frac{\alpha}{4\pi}\right) \left(\frac{\alpha_s}{4\pi}\right)+\beta_1^{(\alpha^2)}\left(\frac{\alpha}{4\pi}\right)^2+\dots\right]\,.
\end{equation}
where the mixed terms, $\beta_1^{(\alpha_s\alpha)}$ and
$\beta_1^{(\alpha\alpha_s)}$, and the pure NLO QED term,
$\beta_1^{(\alpha^2)}$, can be found in
Ref.~\cite{Surguladze:1996hx}. Taking into account a factor four due
to the different definitions of the expansion parameters, one finds:
\begin{equation}\label{eq:NewBetaTerms}
\beta_1^{(\alpha_s\alpha)} = -2\sum_{i=1}^{n_f}
e_q^2\,\qquad\beta_1^{(\alpha\alpha_s)} = -\frac{16}{3}N_c\sum_{i=1}^{n_f} e_q^2\,,\qquad \beta_1^{(\alpha^2)} = -4\left(n_l+N_c\sum_{i=1}^{n_f} e_q^2\right)\,,
\end{equation}
where $N_c=3$ is the number of colours, $e_q$ is the electric charge
of the quark flavour $q$, and $n_f$ and $n_l$ are the number of active
quark and lepton flavours, respectively.

Eq.~(\ref{CoupledEq}) can be written in the vectorial form:
\begin{equation}\label{CoupledEqVect}
\mu^2\frac{\partial {\bm \alpha}}{\partial \mu^2} = {\bm \beta}\left({\bm \alpha}(\mu)\right)\,,
\end{equation}
with:
\begin{equation}
  {\bm \alpha} = {\alpha_s \choose \alpha}\qquad\mbox{and}\qquad  {\bm \beta} = {\beta^{\rm QCD} \choose \beta^{\rm QED}}\,.
\end{equation}
Eq.~(\ref{CoupledEqVect}) is an ordinary differential equation that
can be numerically solved using, for example, Runge-Kutta methods.

The first two terms in eq.~(\ref{eq:NewBetaTerms}) are responsible for
the coupling of the evolution of $\alpha_s$ and $\alpha$, and thus they
introduce a complication that affects both the implementation and the
performance of the code.
One can then ask what is the effect of their
presence and whether their removal makes a substantial difference. 

Fig.~\ref{fig:CouplingEvol} shows the comparison between the
evolution at NLO of both couplings $\alpha_s$ and $\alpha$ including
and excluding the mixed terms in the respective
$\beta$-functions.
The evolution is performed between the $Z$ mass
scale $M_Z$ and 10 TeV with 5 active quark flavours and 3 active
lepton flavours and uses as boundary conditions
$\alpha_s(M_Z) = 0.118$ and $\alpha(M_Z) = 1/128$.
The two curves in
Fig.~\ref{fig:CouplingEvol} are normalised to the respective curves
without mixed terms. It is clear that the mixed terms lead to tiny
relative differences that are at most of $\mathcal{O}(10^{-4})$ at 10
TeV for $\alpha_s$ and $\mathcal{O}(10^{-3})$ at the same scale for
$\alpha$.
Thus it is safe to conclude that the mixed terms in the $\beta$-functions
have a negligible effect on the evolution of the couplings and thus they are 
excluded to simplify the code  and to improve the performance
without introducing any significant inaccuracy.

%%%%%%%%%%%%%%%%%%%%%%%%%%%%%%%%%%%%%%%%%%%%%%%%%%%%%%%%
\begin{figure}[h]
\includegraphics[width=6cm,angle=270]{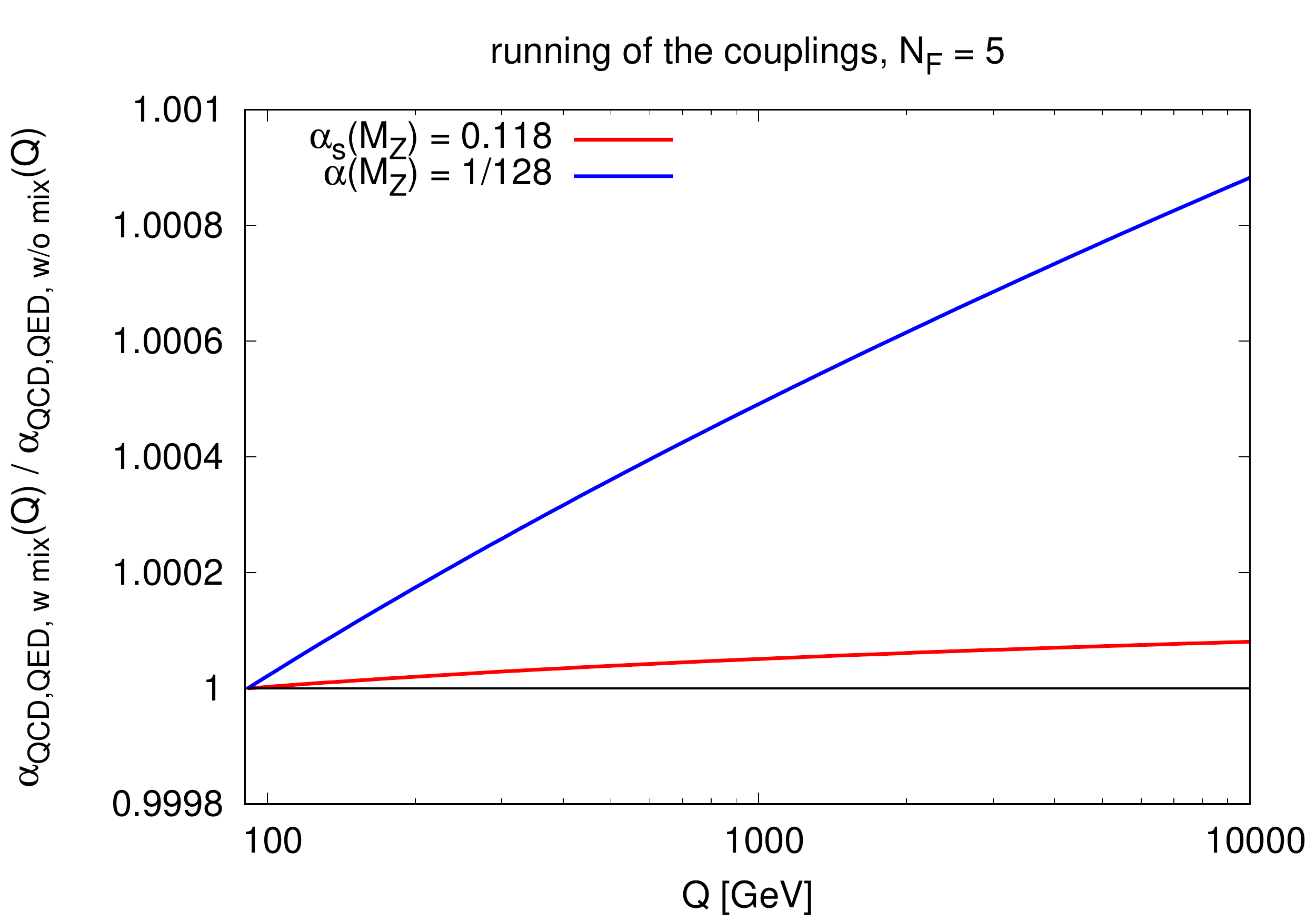} 
\caption{Comparison between the running with the scale
$Q$ of the QCD and QED couplings,
  $\alpha_s$ and $\alpha$, including or not the mixed terms in
  the corresponding $\beta$-functions.
  The curves are normalised to the result of the respective coupling
  running without the mixed terms included in the $\beta$ functions.}
\label{fig:CouplingEvol}
\end{figure}
%%%%%%%%%%%%%%%%%%%%%%%%%%%%%%%%%%%%%%%%%%%%%%%%%%%%%%%%

\subsection{PDF evolution with NLO QED corrections}

Next the implementation of the full NLO QCD+QED corrections to the
DGLAP evolution equations is considered.
The discussion is limited to consideration of
the photon, leptons are not considered.
The first step towards an efficient implementation of the solution of
the DGLAP equations in the presence of QED corrections is the adoption
of a suitable PDF basis that diagonalises the splitting function
matrix, decoupling as many equations as possible.
Such a basis was introduced in the Appendix A of
Ref.~\cite{Bertone:2015lqa} and will be used also here.
Excluding the lepton PDFs, this basis contains 14 independent PDF
combinations and reads:
\begin{equation}\label{eq:EvolBasis}
\begin{array}{ll}
\mbox{\texttt{ 1} : }g & \\
\mbox{\texttt{ 2} : }\gamma & \\
\mbox{\texttt{ 3} : }\displaystyle \Sigma = \Sigma_u + \Sigma_d & \quad
\mbox{\texttt{9} : }\displaystyle V =V_u +  V_d\\
\mbox{\texttt{ 4} : } \displaystyle \Delta_\Sigma = \Sigma_u - \Sigma_d& \quad\displaystyle 
\mbox{\texttt{10} : } \Delta_V = V_u - V_d\\
\mbox{\texttt{ 5} : }T_1^u = u^+ - c^+ &\quad \mbox{\texttt{11} : }V_1^u = u^- - c^- \\
\mbox{\texttt{ 6} : }T_2^u = u^+ + c^+ - 2t^+ &\quad \mbox{\texttt{12} : }V_2^u = u^- + c^- - 2t^-\\
\mbox{\texttt{ 7} : }T_1^d = d^+ - s^+ &\quad \mbox{\texttt{13} : }V_1^d = d^- - s^- \\
\mbox{\texttt{ 8} : }T_2^d = d^+ + s^+ - 2b^+ &\quad \mbox{\texttt{14}
                                               : }V_2^d = d^- + s^- -
                                               2b^-\\
\end{array}
\end{equation}
where $q^\pm = q\pm\overline{q}$ with
$q = u,d,s,c,b,t$. In addition:
\begin{equation}
\begin{array}{ll}
\Sigma_u = u^++c^++t^+, &\quad V_u = u^-+c^-+t^-,\\
\\
\Sigma_d = d^++s^++b^+,&\quad V_d = d^-+s^-+b^-\,.
\end{array}
\end{equation}

The second step is the construction of the splitting function matrix that
determines the evolution of each of the combinations listed in
Eq.~(\ref{eq:EvolBasis}).
To this end, the splitting function matrix $P$ is split into a pure
QCD term $\widetilde{P}$, which only depends on $\alpha_s$, and a
mixed QCD+QED correction term $\overline{P}$, which instead contains
contributions proportional to at least one power of the QED coupling
$\alpha$.
In practice, this means that
\begin{equation}
P = \widetilde{P} + \overline{P}\,,
\end{equation}
where the pure QCD term reads:
\begin{equation}\label{eq:PureQCDSplittings}
\widetilde{P} = \alpha_s \mathcal{P}^{(1,0)} + \alpha_s^2 \mathcal{P}^{(2,0)}+\dots\, ,
\end{equation}
while the term containing the QED coupling is given by:
\begin{equation}\label{eq:QCD+QEDSplittings}
\overline{P} = \alpha \mathcal{P}^{(0,1)} + \alpha_s\alpha \mathcal{P}^{(1,1)}+\alpha^2 \mathcal{P}^{(0,2)} + \dots \, .
\end{equation}
Note that in the r.h.s. of Eqs.~(\ref{eq:PureQCDSplittings})
and~(\ref{eq:QCD+QEDSplittings}) the convention of
Refs.~\cite{deFlorian:2015ujt,deFlorian:2016gvk} is followed to indicate the power
of $\alpha_s$ and $\alpha$ that each splitting function multiplies.

The structure of the pure QCD splitting function matrix
$\widetilde{P}$ as well as the first term in $\overline{P}$, which
represents the pure LO QED correction, were already discussed in
Ref.~\cite{Bertone:2015lqa}.
It is now necessary to analyse the structure of the two additional
terms, namely $\mathcal{P}^{(1,1)}$ and $\mathcal{P}^{(0,2)}$.
Starting with the $\mathcal{O}(\alpha_s\alpha)$ term,
at this perturbative order, the resulting evolution equations read:
\begin{equation}
\begin{array}{rcl}
\displaystyle\left.\mu^2\frac{\partial}{\partial \mu^2}
\begin{pmatrix}
g\\
\gamma\\
\Sigma\\
\Delta_\Sigma
\end{pmatrix}
\right|_{\mathcal{O}(\alpha_s \alpha)} &=& \displaystyle \begin{pmatrix}
e_\Sigma^2 \mathcal{P}^{(1,1)}_{gg}      & e_\Sigma^2 \mathcal{P}^{(1,1)}_{g\gamma} & \eta^+\mathcal{P}^{(1,1)}_{gq} & \eta^-\mathcal{P}^{(1,1)}_{gq} \\
e_\Sigma^2 \mathcal{P}^{(1,1)}_{\gamma g} & e_\Sigma^2 \mathcal{P}^{(1,1)}_{\gamma\gamma} & \eta^+\mathcal{P}^{(1,1)}_{\gamma q} &\eta^-\mathcal{P}^{(1,1)}_{\gamma q} \\
2 e_\Sigma^2 \mathcal{P}^{(1,1)}_{qg}    & 2 e_\Sigma^2 \mathcal{P}^{(1,1)}_{q\gamma} & \eta^+\mathcal{P}^{+(1,1)}  & \eta^-\mathcal{P}^{+(1,1)}\\
2 \delta_e^2 \mathcal{P}^{(1,1)}_{qg} & 2 \delta_e^2 \mathcal{P}^{(1,1)}_{q\gamma} &\eta^-\mathcal{P}^{+(1,1)} &\eta^+\mathcal{P}^{+(1,1)}
\end{pmatrix}\otimes
\begin{pmatrix}
g\\
\gamma\\
\Sigma\\
\Delta_\Sigma
\end{pmatrix}
\end{array}\,,
\end{equation}

\begin{equation}
\displaystyle\left.\mu^2\frac{\partial}{\partial \mu^2}
\begin{pmatrix}
V\\
\Delta_V
\end{pmatrix} \right|_{\mathcal{O}(\alpha_s \alpha)}= 
\begin{pmatrix}
\eta^+\mathcal{P}^{-(1,1)} & \eta^-\mathcal{P}^{-(1,1)} \\
\eta^-\mathcal{P}^{-(1,1)} & \eta^+\mathcal{P}^{-(1,1)} 
\end{pmatrix}\otimes
\begin{pmatrix}
V\\
\Delta_V
\end{pmatrix}\,,
\end{equation}

\begin{equation}
\begin{array}{ll}
\begin{array}{rcl}
\displaystyle \left.\mu^2\frac{\partial T^u_{1,2}}{\partial \mu^2}\right|_{\mathcal{O}(\alpha_s \alpha)} &=&
\displaystyle e_u^2\mathcal{P}^{+(1,1)}\otimes T^u_{1,2}
\end{array}\,, &
\begin{array}{rcl}
\displaystyle \left.\mu^2\frac{\partial T^d_{1,2}}{\partial \mu^2}\right|_{\mathcal{O}(\alpha_s \alpha)} &=&
\displaystyle e_d^2\mathcal{P}^{+(1,1)} \otimes T^d_{1,2}
\end{array}\,,
\\
\\
\begin{array}{rcl}
\displaystyle \left.\mu^2\frac{\partial V^u_{1,2}}{\partial \mu^2}\right|_{\mathcal{O}(\alpha_s \alpha)} &=&
\displaystyle e_u^2\mathcal{P}^{-(1,1)} \otimes V^u_{1,2}
\end{array}\,, &
\begin{array}{rcl}
\displaystyle \left.\mu^2\frac{\partial V^d_{1,2}}{\partial \mu^2}\right|_{\mathcal{O}(\alpha_s \alpha)} &=&
\displaystyle e_d^2\mathcal{P}^{-(1,1)}\otimes V^d_{1,2}
\end{array}\,.
\end{array}
\end{equation}
where $\otimes$ indicates the Mellin convolution and where 
\begin{equation}
\begin{array}{rcl}
e_{\Sigma}^{2}& \equiv &\displaystyle
N_c(n_ue_{u}^{2}+n_de_{d}^{2})\,,\\
\\
\delta_e^2 & \equiv &\displaystyle N_c(n_u e_u^2 -n_d e_d^2)\,,\\
\\
\eta^{\pm} & \equiv & \displaystyle \frac{1}{2}\left(e_{u}^{2}\pm
  e_{d}^{2}\right)\,,\\
\end{array}
\end{equation}
with $e_u$ and $e_d$ the electric charges of the up- and down-type
quarks, and $n_u$ and $n_d$ the number of up- and down-type active
quark flavours (such that $n_u+n_d=n_f$).
%
%Once expressed this basis, it is possible to include the
%$\mathcal{O}(\alpha_s\alpha)$ corrections in the solution of the DGLAP
%QCD+QED equations in APFEL as done in Ref.~\cite{Bertone:2015lqa}.

Next the $\mathcal{O}(\alpha^2)$ corrections as considered.
The expressions of the splitting functions at this order have been
presented in Ref.~\cite{deFlorian:2016gvk}.
There are two relevant new features that distinguish these corrections
from the $\mathcal{O}(\alpha)$ and the $\mathcal{O}(\alpha_s\alpha)$
ones.
The first one is that, contrary to the other cases in which the
electric charges appears to the second power at most, here they appear
up to the fourth power.
As a consequence, new couplings must be introduced:
\begin{equation}
\begin{array}{l}
e_{\Sigma}^4 = N_c(n_{u} e_u^4 + n_{d} e_d^4)\,,\\
\\
\delta_e^4 = N_c(n_{u} e_u^4 - n_{d} e_d^4)\,.
\end{array}
\end{equation}
The second feature is that the dependence on the electric charges of
some of the $\mathcal{O}(\alpha^2)$ splitting functions is not
factorisable as was the case for all the $\mathcal{O}(\alpha)$ and
$\mathcal{O}(\alpha_s\alpha)$ ones and therefore a distinction must be made
 between up- and down-type splitting functions.
Taking into account these features, it is possible to show that the
$\mathcal{O}(\alpha^2)$ contributions to the DGLAP equations take the
following form:
\begin{equation}
%\begin{array}{rcl}
\begin{array}{c}
\displaystyle\left.\mu^2\frac{\partial}{\partial \mu^2}
\begin{pmatrix}
g\\
\gamma\\
\Sigma\\
\Delta_\Sigma
\end{pmatrix}
  \right|_{\mathcal{O}(\alpha^2)} =\\
\\
 \displaystyle \frac12\begin{pmatrix}
    0 & 0 & 0 & 0 \\
    0 & 2e_\Sigma^4 \mathcal{P}_{\gamma\gamma}^{(0,2)} & e_u^4 \mathcal{P}_{\gamma
      u}^{(0,2)} + e_d^4 \mathcal{P}_{\gamma d} & e_u^4 \mathcal{P}_{\gamma u}^{(0,2)} - e_d^4 \mathcal{P}_{\gamma d}^{(0,2)}\\
    0 & 4 e_\Sigma^4 \mathcal{P}^{(0,2)}_{q\gamma} &
    e_u^4\mathcal{P}_{uu}^{+(0,2)}
    +e_d^4\mathcal{P}_{dd}^{+(0,2)}+2\eta^+e_\Sigma^2\mathcal{P}^{S(0,2)}_{qq} & e_u^4\mathcal{P}_{uu}^{+(0,2)}-e_d^4\mathcal{P}_{dd}^{+(0,2)} + 2\eta^-e_\Sigma^2\mathcal{P}^{S(0,2)}_{qq}\\
    0 & 4 \delta_e^4 \mathcal{P}^{(0,2)}_{q\gamma} & e_u^4\mathcal{P}_{uu}^{+(0,2)}
    -e_d^4\mathcal{P}_{dd}^{+(0,2)}+2\eta^-\delta_e^2
    \mathcal{P}^{S(0,2)}_{qq} & e_u^4\mathcal{P}_{uu}^{+(0,2)}+e_d^4\mathcal{P}_{dd}^{+(0,2)} + 2\eta^+\delta_e^2 \mathcal{P}^{S(0,2)}_{qq}
\end{pmatrix}\otimes
\begin{pmatrix}
g\\
\gamma\\
\Sigma\\
\Delta_\Sigma
\end{pmatrix}\,,
\end{array}
\end{equation}

\begin{equation}
\displaystyle\left.\mu^2\frac{\partial}{\partial \mu^2}
\begin{pmatrix}
V\\
\Delta_V
\end{pmatrix} \right|_{\mathcal{O}(\alpha^2)}= \frac12
\begin{pmatrix}
e_u^4\mathcal{P}_{uu}^{-(0,2)}+e_d^4\mathcal{P}_{dd}^{-(0,2)} & e_u^4\mathcal{P}_{uu}^{-(0,2)}-e_d^4\mathcal{P}_{dd}^{-(0,2)} \\
e_u^4\mathcal{P}_{uu}^{-(0,2)}-e_d^4\mathcal{P}_{dd}^{-(0,2)} & e_u^4\mathcal{P}_{uu}^{-(0,2)}+e_d^4\mathcal{P}_{dd}^{-(0,2)} 
\end{pmatrix}\otimes
\begin{pmatrix}
V\\
\Delta_V
\end{pmatrix}\,,
\end{equation}

\begin{equation}
\begin{array}{ll}
\begin{array}{rcl}
\displaystyle \left.\mu^2\frac{\partial T^u_{1,2}}{\partial \mu^2}\right|_{\mathcal{O}(\alpha^2)} &=&
\displaystyle e_u^4\mathcal{P}_{uu}^{+(0,2)}\otimes T^u_{1,2}
\end{array}\,, &
\begin{array}{rcl}
\displaystyle \left.\mu^2\frac{\partial T^d_{1,2}}{\partial \mu^2}\right|_{\mathcal{O}(\alpha^2)} &=&
\displaystyle e_d^4\mathcal{P}_{dd}^{+(0,2)} \otimes T^d_{1,2}
\end{array}\,,
\\
\\
\begin{array}{rcl}
\displaystyle \left.\mu^2\frac{\partial V^u_{1,2}}{\partial \mu^2}\right|_{\mathcal{O}(\alpha^2)} &=&
\displaystyle e_u^4\mathcal{P}_{uu}^{-(0,2)} \otimes V^u_{1,2}
\end{array}\,, &
\begin{array}{rcl}
\displaystyle \left.\mu^2\frac{\partial V^d_{1,2}}{\partial \mu^2}\right|_{\mathcal{O}(\alpha^2)} &=&
\displaystyle e_d^4\mathcal{P}_{dd}^{-(0,2)}\otimes V^d_{1,2}
\end{array}\,.
\end{array}
\end{equation}
It should be noted that, as compared to the expressions for
$\mathcal{P}^{(0,2)}$ presented in Ref.~\cite{deFlorian:2016gvk},
the electric charges haven been factored out in such a way that the
expressions of the splitting functions are either independent from the
electric charges or depend on them only through the ratio
$e_\Sigma^2/e_q^2$.
%
%As before, once the $\mathcal{O}(\alpha^2)$ corrections to the QCD+QED
%combined evolution equations are expressed in this form, it is
%possible to include them in {\tt APFEL} and to solve them using the
%same numerical techniques as for the other cases.

As an illustration, the effects of the
$\mathcal{O}(\alpha_s\alpha)$ and $\mathcal{O}(\alpha^2)$ corrections
to the DGLAP evolution equations on the $\gamma\gamma$ luminosity are quantified at
$\sqrt{s} = 13$ TeV. This luminosity is defined as:
\begin{equation}\label{eq:GammaGammaLumi}
  \mathcal{L}_{\gamma\gamma}(M_X) = \frac1{s}\int_{M_X^2/s}^1
  \frac{dx}{x} \gamma(x,M_X^2) \gamma\left(\frac{M_X^2}{xs},M_X^2\right)\,,
\end{equation}
as a function of the final state invariant mass $M_X$.
Fig.~\ref{fig:GammaGammaLumi} illustrates the behaviour of
$\mathcal{L}_{\gamma\gamma}$ computed using the photon PDF from the
NNPDF30QED NLO set as an input at $Q_0 = 1$ GeV and evolved to $Q=M_X$
including, on top of the pure QCD NLO evolution, the following
corrections:
\begin{itemize}
\item the $\mathcal{O}(\alpha)$ corrections only,
\item same as above, adding also the mixed
  $\mathcal{O}(\alpha_s\alpha)$ corrections, and
\item the complete NLO QCD+QED corrections accounting for the
  $\mathcal{O}(\alpha+\alpha_s\alpha+\alpha^2)$ effects.
\end{itemize}
The results are shown normalised to the predictions obtained with LO
QED corrections only.
It is clear that the $\mathcal{O}(\alpha_s\alpha)$ and
$\mathcal{O}(\alpha^2)$ corrections have a small but non-negligible
impact on the $\gamma\gamma$-luminosity. In particular, these
corrections suppress $\mathcal{L}_{\gamma\gamma}$ by around 10\% at
relatively small values of $M_X$, while the suppression gradually
shrinks to 1-2\% as $M_X$ increases. As expected, most of this effect
comes from the $\mathcal{O}(\alpha_s\alpha)$ corrections, while the
impact of the $\mathcal{O}(\alpha^2)$ ones is substantially smaller.
The $\mathcal{O}(\alpha_s\alpha)$ and $\mathcal{O}(\alpha^2)$
corrections to the DGLAP evolution have more recently been implemented
in the {\tt QEDEVOL} package~\cite{Sadykov:2014aua} based on the
{\tt QCDNUM} evolution code~\cite{Botje:2010ay}. {\tt APFEL} and {\tt
  QEDEVOL} have been found to be in excellent agreement.

%%%%%%%%%%%%%%%%%%%%%%%%%%%%%%%%%%%%%%%%%%%%%%%%%%%%%%%%
\begin{figure}[t]
\includegraphics[width=10cm]{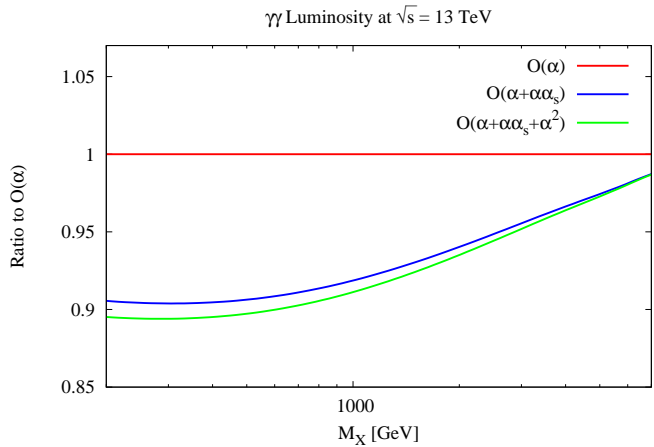} 
\caption{The photon-photon PDF luminosity $\mathcal{L}_{\gamma\gamma}$ at $\sqrt{s} = 13$ TeV as a
  function of the final state invariant mass $M_X$.
  The results with the photon evolved
  with only the $\mathcal{O}(\alpha)$ corrections
  are compared with the corresponding results taking into account the 
  $\mathcal{O}(\alpha+\alpha_s\alpha)$ corrections
  and the complete
  $\mathcal{O}(\alpha+\alpha_s\alpha+\alpha^2)$ effects,
  normalised in all three cases to the $\mathcal{O}(\alpha)$ result.
  The calculation has been performed using the central value of the NNPDF3.0QED NLO
  fit.  }
\label{fig:GammaGammaLumi}
\end{figure}
%%%%%%%%%%%%%%%%%%%%%%%%%%%%%%%%%%%%%%%%%%%%%%%%%%%%%%%%

\subsection{DIS structure functions}

When considering NLO QCD+QED corrections to the DIS structure
functions, it becomes necessary to include into the hard cross
sections all the $\mathcal{O}(\alpha)$ diagrams where one photon is
either in the initial state or emitted from an incoming quark (or
possibly an incoming lepton).
Such diagrams, being of purely QED origin, have associated coefficient
functions that can be easily derived from the QCD expressions by
properly adjusting the colour factors.
This correspondence holds irrespective of whether mass effects are
included.

The main complication of the inclusion of these corrections arises
from their flavour structure. In fact, in the case of
quarks the isospin symmetry is broken due to the fact that the
coupling of the photon is proportional to the squared charge of the
parton to which it couples (a quark or a lepton).
In the following, the neutral-current (NC)
case, where lepton and proton exchange a neutral boson $\gamma^*/Z$,
and the charged-current (CC) case, where instead lepton and proton
exchange a charged $W$ boson, are addressed separately.

First  the $\mathcal{O}(\alpha)$
contributions to a generic NC structure function $F$ are considered.
Due to the fact that to this order there is no mixing between QCD and
QED, such corrections can easily be derived from the
$\mathcal{O}(\alpha_s)$ coefficient functions just by adjusting the
colour factors by setting $C_F=T_R=1$ and $C_A=0$.
Referring, $e.g.$, to the expressions reported in
Ref.~\cite{Ellis:1991qj}, the coefficient functions become:
\begin{equation}\label{eq:alphaCFs}
\begin{array}{rcl}
\displaystyle C_{i;q}^{(\alpha)} &=& \displaystyle \frac{C_{i;q}^{(\alpha_s)}}{C_F}\\
\\
\displaystyle C_{i;\gamma}^{(\alpha)} &=& \displaystyle \frac{C_{i;g}^{(\alpha_s)}}{T_R}
\end{array}\qquad i = 2,L,3\,.
\end{equation}
In order to construct the corresponding structure
functions, considering that the coupling between a photon and a quark
of flavour $q$ is proportional to $e_q^2$, one also needs to adjust
the electroweak couplings should be adjusted as follows:
 \begin{equation}
\begin{array}{rcl}
\widetilde{B}_q &=& B_qe_q^2\quad\mbox{for}\quad F_2,F_L\,, \\
\\
\widetilde{D}_q &=& D_qe_q^2\quad\mbox{for}\quad F_3\,, \\
\end{array}
\end{equation}
where $B_q$ and $D_q$ are defined, $e.g.$, in
Ref.~\cite{Adloff:2003uh}.
Following this prescription, it is possible to write the
$\mathcal{O}(\alpha)$ contributions to the NC structure functions as:
\begin{equation}
\begin{array}{rcl}
F_{2,L}^{{\rm NC},(\alpha)} &=& \displaystyle x \sum_{q} \widetilde{B}_q\left[C_{2,L;q}^{(\alpha)}\otimes
(q+\overline{q}) + C_{2,L;\gamma}^{(\alpha)} \otimes \gamma
                         \right]\,,\\
\\
xF_3^{{\rm NC},(\alpha)} &=& \displaystyle x \sum_{q} \widetilde{D}_q\left[C_{3;q}^{(\alpha)}\otimes
(q-\overline{q}) + C_{3;\gamma}^{(\alpha)} \otimes \gamma
                         \right]\,.
\end{array}
\end{equation}
This structure holds for both massless and massive
structure functions. This aspect is relevant to the construction of
the FONLL general-mass structure functions.

For the CC case the procedure to obtain the
expressions of the $\mathcal{O}(\alpha)$ coefficient functions is
exactly the same as in the NC case (see
Eq.~(\ref{eq:alphaCFs})).
However, this case is more complicated
because the flavour structure of CC structure functions is more
complex.
Taking into account the presence of a factor $e_q^2$ every time that a
quark of flavour $q$ couples to a photon, the $\mathcal{O}(\alpha)$
corrections to the CC structure functions $F_2$ and $F_L$ for the
production of a neutrino or an anti-neutrino take the form:
\begin{equation}\label{compactNu}
\begin{array}{rcl}
F_{2,L}^{{\rm CC},\nu,(\alpha)} &=& \displaystyle
                              x\sum_{U=u,c,t}\sum_{D=d,s,b}|V_{UD}|^2\left[C_{2,L;q}^{(\alpha)}\otimes\left(e_D^2D +e_U^2\overline{U}\right) +2 C_{2,L;\gamma}^{(\alpha)}\otimes\gamma\right]\,,\\
\\
F_{2,L}^{{\rm CC},\overline{\nu},(\alpha)} &=& \displaystyle
x\sum_{U=u,c,t}\sum_{D=d,s,b}|V_{UD}|^2\left[C_{2,L;q}^{(\alpha)}\otimes\left(e_D^2\overline{D}
    +e_U^2U\right) +2 C_{2,L;\gamma}^{(\alpha)}\otimes\gamma\right]\,,
\end{array}
\end{equation}
where $V_{UD}$ are the elements of the CKM matrix.
The flavour structure of $F_3$ is instead slightly different:
\begin{equation}\label{compactNuF3}
\begin{array}{rcl}
xF_3^{{\rm CC},\nu,(\alpha)} &=& \displaystyle
                              x\sum_{U=u,c,t}\sum_{D=d,s,b}|V_{UD}|^2\left[C_{3;q}^{(\alpha)}\otimes\left(e_D^2D -e_U^2\overline{U}\right) +2 C_{3;\gamma}^{(\alpha)}\otimes\gamma\right]\,,\\
\\
xF_3^{{\rm CC},\overline{\nu},(\alpha)} &=& \displaystyle
x\sum_{U=u,c,t}\sum_{D=d,s,b}|V_{UD}|^2\left[C_{3;q}^{(\alpha)}\otimes\left(-e_D^2\overline{D}
    +e_U^2U\right) +2 C_{3;\gamma}^{(\alpha)}\otimes\gamma\right]\,.
\end{array}
\end{equation}
In order to simplify the implementation, it is advantageous to assume
that, in these particular corrections, the CKM matrix is a $3 \times 3$
unitary matrix. Note however that the exact CKM matrix is still used
in the QCD part of the structure functions.
This approximation introduces an inaccuracy of the order of the QED
coupling $\alpha$ times the value of the off-diagonal elements of the
CKM matrix and therefore it is numerically negligible.

As an illustration of the impact of the $\mathcal{O}(\alpha)$
correction on the DIS structure functions, Fig.~\ref{fig:StructFuncs}
shows the effect of introducing these contributions on top of the pure
QCD computation at NLO.
The plots are produced by evolving the
NNPDF3.0QED NLO set from $Q_0=1$ GeV to $Q=100$ GeV including the full
NLO QCD+QED corrections discussed in the previous section and using
the resulting evolved PDFs to compute the NC (left panel) and the CC
(right panel) DIS structure functions in the FONLL-B scheme, including
the $\mathcal{O}(\alpha)$ corrections to the coefficient functions
discussed above.
The predictions are shown normalised to the pure QCD computation where
the QED corrections are absent both in the evolution and in the
computation of the structure functions.

%%%%%%%%%%%%%%%%%%%%%%%%%%%%%%%%%%%%%%%%%%%%%%%%%%%%%%%%
\begin{figure}[t]
\includegraphics[width=6cm,angle=270]{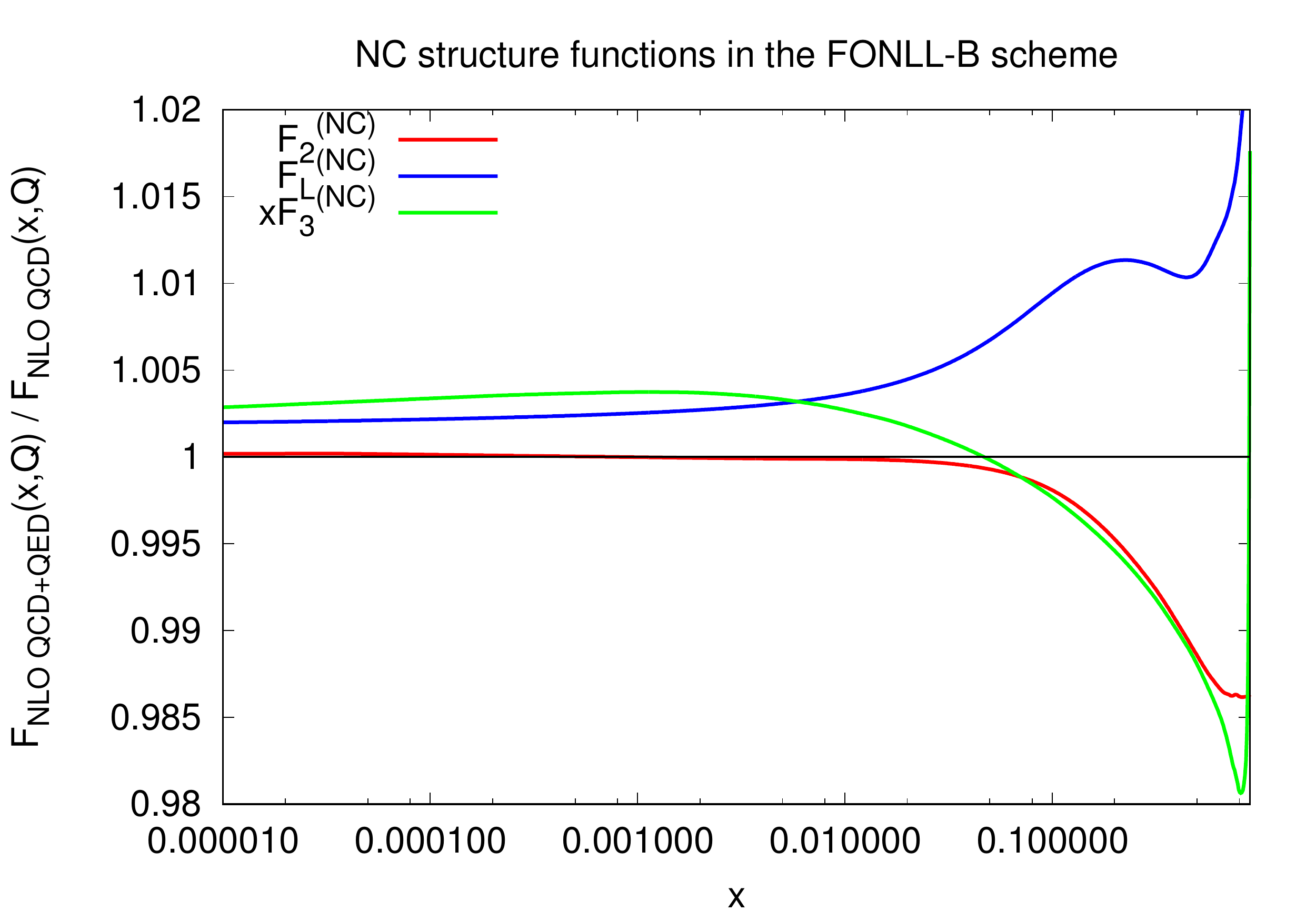}
\includegraphics[width=6cm,angle=270]{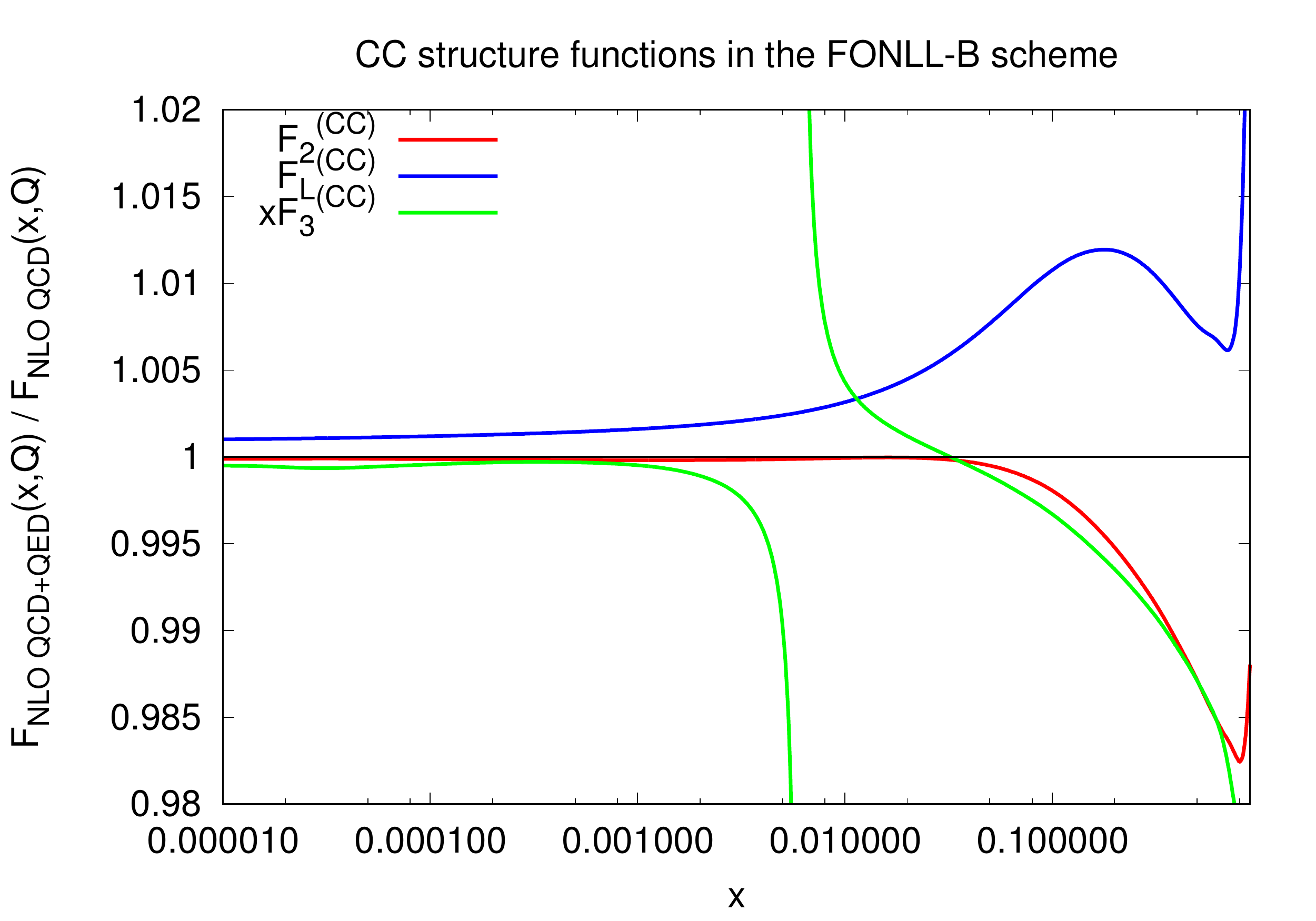}
\caption{The effects of the NLO QED corrections on the neutral-current
(left) and charged-current (right) DIS structure functions
$F_2, F_L$ and $xF_3$, normalised to the pure QCD results.
The calculation has been performed in the FONLL-B general-mass scheme using the
central NNPDF3.0QED NLO
set as input.
Note that QED effects enter both via DGLAP evolution and the
$\mathcal{O}(\alpha)$ DIS coefficient functions.
The behaviour of $xF_3$ in the right plot for $x\sim 0.007$ is explained
by the fact that this structure function exhibits a node in that region.
}
\label{fig:StructFuncs}
\end{figure}
%%%%%%%%%%%%%%%%%%%%%%%%%%%%%%%%%%%%%%%%%%%%%%%%%%%%%%%%

It is clear that the impact of the full NLO QCD+QED corrections is
pretty small especially in the low-$x$ region where it is well below
1\%.
In the large-$x$ region, instead, the presence of a
photon-initiated contribution has a more significant effect because of
the suppression of the QCD distributions (quarks and gluon) relative
to the photon PDF and the impact of the QED corrections reaches the
2\% level.
It should be stressed that the behaviour around
$x=10^{-2}$ of the CC $xF_3$ (green curve in the right panel) is driven
by a change of sign of the predictions (in other words,
$xF_3$ exhibits a node in this region) so that the ratio diverges.

%%%%%%%%%%%%%%%%%%%%%

\bibliographystyle{JHEP}

\bibliography{main}

%%%%%%%%%%%%%%%%%%%%%%%%%%%%%%%%%%%%%%%%%%%%%%%%%%

\end{document}